\documentclass{aa} 
\usepackage{graphics}
\usepackage{times}
\usepackage{amssymb}
\newcommand{\diff}{\mathrm d}
\newcommand{\imag}{\mathrm i}
\renewcommand{\O}{\mathcal O}
\newcommand{\Cov}{\mathrm{Cov}}
\newcommand{\sgn}{\mathop{\mathrm{sgn}}\nolimits}
\newcommand{\CL}{\mathit{CL}}
\newcommand{\sint}[2]{\int_{\hbox to 0pt{$\scriptstyle
#1$\hss}}^{\hbox to 0pt{$\scriptstyle #2$\hss}}}
\begin{document}

   \thesaurus{02(12.03.3;  
                 12.07.1;  
                 12.04.1;  
                 11.03.1;  
                 11.04.1)} 
   \title{Weak lensing and cosmology}
   \titlerunning{Weak lensing \& cosmology}
   \author{Marco Lombardi \and Giuseppe Bertin}
   \authorrunning{M. Lombardi \& G. Bertin}
   \offprints{M. Lombardi}
   \mail{lombardi@sns.it}
   \institute{Scuola Normale Superiore,
              Piazza dei Cavalieri 7, I 56126 Pisa, Italy}
   \date{Received ***date***; accepted ***date***}

   \maketitle

   \begin{abstract}

   Recently, it has been shown that it is possible to reconstruct the
   projected mass distribution of a cluster from weak lensing provided
   that both the geometry of the universe and the probability
   distribution of galaxy redshifts are known; actually, when
   additional photometric data are taken to be available, the galaxy
   redshift distribution could be determined jointly with the cluster
   mass from the weak lensing analysis. In this paper we develop, in
   the spirit of a ``thought experiment,'' a method to constrain the
   geometry of the universe from weak lensing, provided that the
   redshifts of the source galaxies are measured. The quantitative
   limits and merits of the method are discussed analytically and with
   a set of simulations, in relation to point estimation, interval
   estimation, and test of hypotheses for homogeneous
   Friedmann-Lema\^\i tre models. The constraints turn out to be
   significant when a few thousand source galaxies are used.

   \end{abstract}

%

\section{Introduction}

It has long been recognized (see Tyson {\it et al}.\ 1984) that the
gravitational field of a cluster acts on the images of distant
galaxies by changing their orientation so that their major axes tend
to become perpendicular to the direction of the center of the
cluster. In turn (see Kaiser \& Squires 1993) the mean ellipticity of
the galaxy images can be used to measure the shear of the lens, a
quantity related to the two-dimensional mass distribution (projected
along the line of sight). By now a number of reconstruction techniques
have been proposed and tested (see, e.g., Kaiser {\it et al}.\ 1995;
Seitz \& Schneider 1996; Hoekstra {\it et al}. 1998).

In the reconstruction process, information on the ratio of the
distances from the observer to the source galaxies and from the lens
to the galaxies is used. In practice, in the limit where the galaxies
are much farther than the lens, the reconstruction process becomes
independent of the distance of the source galaxies (Schneider \& Seitz
1995, Seitz \& Schneider 1995). This, of course, is very useful, as
usually the redshifts of the sources are not available. If this limit
cannot be taken (as it happens if the cluster itself has a relatively
high redshift, e.g., $z_\mathrm d \gtrsim 0.5$), then one has to take
into account the different redshifts of the background galaxies. The
result of the gravitational lensing {\it does depend\/} on the
geometry of the universe.

Seitz \& Schneider (1997) have shown that it is possible to
reconstruct the mass distribution of the lens provided that both the
probability distribution of galaxy redshifts and the geometry of the
universe are known (e.g., the values of $\Omega$ and $\Omega_\Lambda$
in the standard Friedmann-Lema\^\i tre cosmology). If other data (in
particular, the angular sizes and luminosities of the source galaxies)
are taken to be available, then the problem of the joint determination
of the galaxy redshift distribution and of the lens mass density, for
an assumed geometry of the universe, can be solved using the so called
``lens parallax method'' (Bartelmann \& Narayan 1995). This latter
method is based on the decrease of the surface brightness of galaxies
and on the corresponding increase of the lens strength with
redshift. A third possibility, of determining the properties of the
lens and the geometrical characteristics of the universe from an
assumed knowledge of the distances to the source galaxies, is not
natural because of the prohibitive demands posed by the measurement of
a large number of redshifts. However, this last option is actually
most natural {\it in principle}, given the fact that redshifts are the
quantities that can be best measured {\it directly}.

In recent years the technique to obtain photometric redshifts (Baum
1962) has been greatly improved and tested. A large number of
photometric redshifts of very faint sources ($m_B > 25$) can be
obtained without additional telescope time (with respect to
imaging). For the Hubble Deep Field more than one thousand photometric
redshifts have been estimated (Lanzetta {\it et al}.\ 1996). Moreover,
the technique can be rather accurate (provided that a good set of
filters is used). For example, in a first test on a sample of 27
galaxies in the Hubble Deep Field, by comparison with the
spectroscopic redshifts now available, more than $68\%$ of the
photometric redshifts have errors $| \Delta z | < 0.1$, and all
redshifts have errors $| \Delta z | < 0.3$ (Hogg {\it et
al}. 1998). This suggests that in a not far future we might use
photometric redshift information for galaxies lensed by a cluster.

In this paper we show how we can determine both the lens mass
distribution and the geometry of the universe if the redshifts of the
source galaxies are known. The method outlined here could be used, in
a Friedmann-Lema\^\i tre universe, to get information on the density
parameter $\Omega$ and on the cosmological constant parameter
$\Omega_\Lambda$, or could be used to test different cosmologies. At
this stage the study proposed here may be seen as a ``thought
experiment'' because of the very high number of redshifts that are
found to be required in order to constrain the geometry significantly.
We will show below that the galaxies that one should observe can
hardly be found behind a single cluster at the magnitudes currently
attainable. However, it is possible to combine data from different
clusters. Combining the study of $5$--$10$ clusters would lead to
statistically meaningful constraints. With such a device, the
application of the present method might even become feasible in a not
too far future. Surprisingly, simulations show that the method applied
to observations considered to be realistic for current plans of the
Next Generation Space Telescope (NGST) should provide significant
constraints on the cosmological parameters even if based on a single
cluster.

The paper is organized as follows. In Sect. 2 we recall the main
lensing equations generalized to sources at different distances and
introduce the ``cosmological weight,'' a quantity that gives the
strength of the lens scaled to the relevant distances involved. The
reconstruction of the cluster mass distribution and of the
cosmological weight is described in Sect.~3. Section~4 addresses the
issue of the invariance properties associated with the reconstruction
analysis. Section~5 introduces the determination of the cosmological
parameters by recognizing the limits on the measurement of the
cosmological weight. The results of a wide set of simulations are
described in detail in Sect.~6. Section~7 addresses the problem of the
actual feasibility of a measurement based on the procedure outlined in
the paper. Finally, in Sect.~8, we summarize the main results
obtained. Four appendices contain detailed discussions and derivations
of some important results.


\section{Distance dependent lensing relations}

This section is aimed at defining the basic mathematical framework and
mostly follows the article by Seitz \& Schneider (1997). Let us
consider a lens at redshift $z_\mathrm d$ with two-dimensional
projected mass distribution $\Sigma(\vec\theta)$, where $\vec\theta$
is a two-dimensional vector representing a direction on the sky (the
region of interest on the sky is very small and can be considered
flat). For a source at redshift $z$ we define the critical density
$\Sigma_\mathrm c(z)$ as
\begin{equation}
  \Sigma_\mathrm c (z) = \cases{ 
  \infty & for $z \le z_\mathrm d \; ,$ \cr
  \displaystyle\frac{c^2 D(z)}{4 \pi G D(z_\mathrm d) D(z_\mathrm d,
  z)} & otherwise$\; .$ \cr}
\end{equation}
Here $D(z)$ and $D(z_\mathrm d, z)$ are, respectively, the angular
diameter-distance from the observer to an object at redshift $z$, and
from the lens to the same object. The quantities $D(z_\mathrm d)$,
$D(z)$, and $D(z_\mathrm d, z)$ replace the Euclidean distances,
respectively $D_\mathrm{od}$, $D_\mathrm{os}$, and $D_\mathrm{ds}$,
used in Paper~I (Lombardi \& Bertin 1998a).

It is useful to define a redshift independent critical density as
\begin{equation}
  \Sigma_\mathrm c^\infty = \lim_{z \rightarrow \infty} \Sigma_\mathrm
  c (z) \; ,
  \label{Sigma_c}
\end{equation}
related to the redshift dependent critical density through a
``cosmological weight'' function
\begin{equation}
  w(z) = \frac{\Sigma_\mathrm c^\infty}{\Sigma_\mathrm c(z)} \propto
  \frac{D(z_\mathrm d, z)}{D(z)} \; .
  \label{w(z)}
\end{equation}
In principle, the above relations could be the starting point also for
investigations based on non-standard cosmological models, using the
constraints on the cosmological weight $w(z)$ provided by the
gravitational lensing analysis to be described below. In practice, for
the rest of the paper we will focus on the standard Friedmann-Lema\^\i
tre cosmology, where the angular diameter-distance $D(z)$ can be
written as (Peebles 1993; see also Kayser {\it et al}.\ 1997)
\begin{eqnarray}
  && D(z) = D(0, z) \; , \\
  && D(z_\mathrm d, z) = \frac{c}{H_0 (1 + z) \sqrt{\Omega_R}} \sinh
  \biggl[ \sqrt{\Omega_R} \int_{z_\mathrm d}^z \frac{\diff z'}{E(z')}
  \biggr] \; , \label{D(z)}
\end{eqnarray}
where
\begin{equation}
  E(z') = \sqrt{\Omega (1+z')^3 + \Omega_R (1+z')^2 + \Omega_\Lambda}
  \; ,
\end{equation}
and, we recall, $\Omega + \Omega_\Lambda + \Omega_R = 1$.  The
hyperbolic sine in Eq.~(\ref{D(z)}) is characteristic of an open
universe ($\Omega_R > 0$). 

\begin{figure}[!t]
  \resizebox{\hsize}{!}{\includegraphics{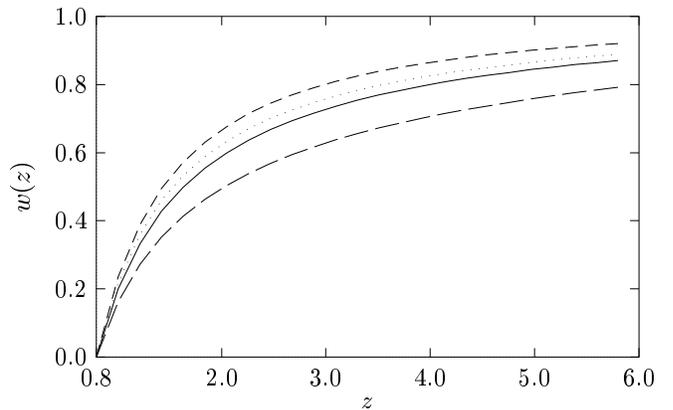}} \caption{The
  weight function for a lens at $z_\mathrm d = 0.8$ in four different
  universes. From the top to the bottom: $\Omega = 0.3$,
  $\Omega_\Lambda = 0$ (dashed); $\Omega = 0.3$, $\Omega_\Lambda =
  0.7$ (dotted); $\Omega = 1$, $\Omega_\Lambda = 0$ (solid); $\Omega =
  1$, $\Omega_\Lambda = 1$ (long dashed).}
\end{figure}

In general, $w(z) \rightarrow 1$ for $z \rightarrow \infty$. As we
will see shortly, the lensing equations depend on the geometry via the
cosmological weight, and thus this is the function that can be
determined from lensing observations.  The function depends in a
complicated manner on the two cosmological parameters $\Omega$ and
$\Omega_\Lambda$. For an Einstein-de Sitter universe ($\Omega=1$,
$\Omega_\Lambda = 0$, so that $\Omega_R = 0$) one has, for $z >
z_\mathrm d$,
\begin{eqnarray}
  && \frac{H_0 D(z_\mathrm d, z)}{c} = \frac{2}{1+z} \left(
  \frac{1}{\sqrt{1 + z_\mathrm d}} - \frac{1}{\sqrt{1 + z}} \right) \;
  , \\
  && w(z) = \frac{\sqrt{1 + z} - \sqrt{1 + z_\mathrm d}}{\sqrt{1 + z}
  - 1} \; . \label{EdS}
\end{eqnarray}
Some examples of weight functions for a lens at redshift $z_\mathrm d
= 0.8$ are given in Fig.~1. It should be noted that the cosmological
weights for ``reasonable'' universes lie very close to each
other. This of course makes the determination of the cosmological
parameters from the measurement of the cosmological weight a difficult
task.

For sources at redshift $z$, we define the Jacobian matrix of the
ray-tracing function as (see Eq.~(3) of Paper~I)
\begin{eqnarray}
  A(\vec\theta, z) &=& \left( \frac{\partial \vec\theta^\mathrm s}{
  \partial \vec\theta} \right) \nonumber\\ &=& \left( \matrix{ 1 -
  w(z) ( \kappa - \gamma_1 ) & w(z) \gamma_2 \cr w(z) \gamma_2 & 1 -
  w(z) ( \kappa + \gamma_1 ) } \right) \; .
  \label{A}
\end{eqnarray}
Here $\kappa = \kappa(\vec\theta) = \Sigma(\vec\theta) /
\Sigma_\mathrm c^\infty$ is the lens density in units of the redshift
independent critical density defined in Eq.~(\ref{Sigma_c}), while
$\gamma = \gamma(\vec\theta)$ is the complex shear. Then the redshift
dependent amplification $\mu(\vec\theta, z)$ is given by
\begin{eqnarray}
  \frac{1}{\mu(\vec\theta, z)} &=& \bigl| \det A(\vec\theta, z) \bigr|
  \nonumber \\
  &=& \bigl[ 1 - w(z) \kappa(\vec\theta) \bigr]^2 - \bigl| w(z)
  \gamma(\vec\theta) \bigr|^2 \; .
  \label{mu(z)}
\end{eqnarray}
This expression gives the magnification produced by the lens on a source
located at redshift $z$. If this quantity can be measured (for
example by using galaxy counts), the mass density of the lens can be
constrained strongly, and an otherwise present invariance property
(see Sect.~4) can be broken.

For a galaxy at redshift $z$ seen at position $\vec{\theta}$ a complex
quantity $\epsilon$ is defined in terms of the galaxy quadrupole
moment $Q_{ij}$ (see Schneider \& Seitz 1995 for a detailed
discussion):
\begin{equation}
  \epsilon = \frac{Q_{11} - Q_{22} + 2\imag Q_{12}}{Q_{11} + Q_{22} +
  2 \sqrt{Q_{11} Q_{22} - Q_{12}^2}} \; .
\end{equation}
At variance with Paper~I and Paper~II (Lombardi \& Bertin 1998b), it
is convenient to measure the ellipticity using this quantity instead
of $\chi$ (see Eq.~(7) of Paper~I), because of the simpler
transformation properties of $\epsilon$ (see Eq.~(\ref{<epsilon>-s})
below). In fact, as shown by Seitz \& Schneider (1997), the observed
ellipticity is related to the source ellipticity through the relation
(the superscript stands for ``source'')
\begin{equation}
  \epsilon = \cases{ 
  \displaystyle \frac{\epsilon^\mathrm s - g(\vec\theta, z)}{1 -
  g^*(\vec\theta, z) \epsilon^\mathrm s} & for $\bigl| g(\vec\theta,z)
  \bigr| \le 1 \; ,$ \cr
  \displaystyle \frac{1 - g(\vec\theta, z) \epsilon^{\mathrm
  s *}}{\epsilon^{\mathrm s*} - g^*(\vec\theta, z)} & otherwise$\; ,$ \cr}
  \label{epsilon-s}
\end{equation}
where the reduced shear $g(\vec\theta, z)$ is defined as
\begin{equation}
  g(\vec\theta, z) = \frac{w(z) \gamma(\vec\theta)}{1 -
  w(z) \kappa(\vec\theta)} \; .
  \label{g}
\end{equation}

From the observation of a large number of galaxies near $\vec\theta$,
with redshifts close to $z$, we can thus determine the reduced shear
$g(\vec\theta, z)$. In fact, it can be shown that the expected mean value
of the observed ellipticities is given by
\begin{equation}
  \langle \epsilon \rangle (\vec\theta, z) = \cases{
  -g(\vec\theta, z) & if $\bigl| g(\vec\theta, z) \bigr| < 1 \; ,$ \cr
  -\frac{\displaystyle 1}{\displaystyle g^*(\vec\theta, z)}
  & otherwise$\; ,$\cr}
  \label{<epsilon>-s}
\end{equation}
under the isotropy hypothesis $\langle \epsilon^\mathrm s \rangle = 0$
for the source galaxy distribution. This expression depends on the
redshift of the galaxies used. Therefore, if we measure the mean value
of the ellipticities for galaxies at different redshifts, we can
obtain information on the cosmological weight function $w(z)$. Once
the cosmological weight function $w(z)$ has been determined, the
cluster mass reconstruction can be performed using an iterative
procedure on Eq.~(\ref{g}), seeded by $\kappa = 0$, and some kernel
analysis (see Paper~II). As shown by Seitz \& Schneider (1997), this
can be carried out even if one does not have the redshift of each
galaxy, provided one knows the galaxy probability distribution in $z$.

In the rest of the paper we assume that the ellipticity dispersion of
the source galaxies is independent of redshift (see also App.~C).


\section{Joint determination of the shear and of the cosmological weight}

Let us now address the goal of this paper, the joint
determination of the lens mass and of the geometry of the universe. In
the following, for simplicity, we will consider a cluster sub-critical
for all $z$, i.e.\ a cluster that cannot produce multiple images,
characterized by $\bigr| g(\vec\theta, z) \bigr| < 1$ for all
$\vec\theta$ and all $z$ (from Eq.~(\ref{mu(z)}), by recalling that
subcritical lenses are also characterized by $\det A > 0$, it can be
shown that this happens when $\kappa(\vec\theta) + \bigl|
\gamma(\vec\theta) \bigr| < 1$ for every $\vec\theta$). General
clusters could be treated basically in the same way, but with
considerably heavier notation. Let the positions $\vec\theta^{(n)}$,
the observed ellipticities $\epsilon^{(n)}$, and the redshifts
$z^{(n)}$ of $N$ source galaxies ($n = 1, \dots, N$) be known,
together with the redshift $z_\mathrm d$ of the cluster.

In the following we will use subscript $0$ for {\it true\/}
quantities. For example, $\kappa_0$ will be the true dimensionless
surface density of the lens, and $w_0(z)$ the true cosmological weight
(given by Eq.~(\ref{w(z)})). For the complex quantities $\gamma$ and
$g$ we will use the notation $\gamma_{0i}$ and $g_{0i}$ to denote the
$i$-th component of the true value. Following the notation used in
statistics, we will use hats for {\it measured\/} quantities
(``estimators''). Thus $\hat\kappa(\vec\theta)$ will be the measured
mass density, and $\hat w(z)$ the measured cosmological weight.

In order to describe the reconstruction process, we start by
considering the weak lensing limit, and then we generalize the results
obtained.

\subsection{Weak lensing limit}

The weak lensing limit is characterized by $\bigl|
\gamma_{0}(\vec\theta) \bigr|, \kappa_0(\vec\theta) \ll 1$, so that
the reduced shear $g_0(\vec\theta, z)$ can be identified with the
product $w_0(z) \gamma_0(\vec\theta)$. In this limit
Eqs.~(\ref{epsilon-s}) and (\ref{<epsilon>-s}) become
\begin{equation}
  \epsilon = \epsilon^\mathrm s - \gamma_0(\vec\theta) w_0(z)
  \label{epsilon}
\end{equation}
and
\begin{equation}
  \langle \epsilon \rangle(\vec\theta, z) = -\gamma_0(\vec\theta) w_0(z)
  \; .
  \label{<epsilon>}
\end{equation}
These relations allow us to estimate, from a set of measurements
$\bigl\{ \epsilon^{(n)} \bigr\}$, one of the two quantities, either
the shear or the cosmological weight function, provided the other is
known. The basic idea is to average the observed ellipticities of
galaxies close to each other either in $\vec\theta$-space or in
redshift, following the general theory of maximum likelihood methods
to minimize errors (e.g., see Eadie {\it et al}., 1971).

If we take $w(z)$ to be known, we can introduce the {\it spatial\/}
weight function $W(\vec\theta, \vec\theta')$ with the property that
$W(\vec\theta, \vec\theta')$ is significantly different from zero only
for $\| \vec\theta - \vec\theta' \|$ small (see Kaiser \& Squires 1993;
Paper II). Then, from Eq.~(\ref{epsilon}), the shear
$\gamma(\vec\theta)$ can be estimated by
\begin{equation}
  \hat\gamma(\vec\theta) = - \frac{\sum_{n=1}^N W\bigl( \vec\theta,
  \vec\theta^{(n)} \bigr) w\bigl( z^{(n)} \bigr)
  \epsilon^{(n)}}{\sum_{n=1}^N W\bigl( \vec\theta, \vec\theta^{(n)}
  \bigr) \bigl[ w\bigl( z^{(n)} \bigr) \bigr]^2} \; .  \label{gamma}
\end{equation}
Here, we recall, the cosmological weight function is taken to be
known.

Similarly, if we take the shear $\gamma(\vec\theta)$ to be available,
we can introduce a {\it redshift\/} weight function $W_z(z, z')$, with
the condition that $W_z(z, z')$ is significantly different from zero
only for $| z - z' |$ small. Then the cosmological weight function
$w(z)$ can be estimated by
\begin{equation}
  \hat w(z) = - \frac{\sum_{n=1}^N W_z \bigl(z, z^{(n)} \bigr) \Re
  \bigl[ \gamma\bigl( \vec\theta^{(n)} \bigr) \epsilon^{(n)*} \bigr]}{
  \sum_{n=1}^N W_z\bigl(z, z^{(n)} \bigr) \bigl| \gamma\bigl(
  \vec\theta^{(n)} \bigr) \bigr|^2} \; , \label{w}
\end{equation}
where $\Re$ and the asterisk denote real part and complex conjugate
(we recall that $\epsilon$ and $\gamma$ are complex quantities).

The two equations can be used to obtain by {\it iteration\/} the joint
determination of the shear and of the cosmological weight function. We
start with a guess for the weight function $w(z)$ (e.g., the
expression known for the Einstein-de~Sitter cosmology given in
Eq.~(\ref{EdS})), and calculate the shear using
Eq.~(\ref{gamma}). Then we insert the estimated shear
$\hat\gamma(\vec\theta)$ in Eq.~(\ref{w}) and thus obtain a first
estimate for $w(z)$. We then go back to Eq.~(\ref{gamma}) and obtain a
new determination of $\hat\gamma(\vec\theta)$, etc.

As a by-product of the analysis, by means of standard methods
(Seitz \& Schneider 1996; Paper~II), we can also reconstruct, from the
estimate $\hat\gamma(\vec\theta)$, the dimensionless mass distribution
$\kappa(\vec\theta)$.

At the end of the iteration process, it can be shown that the
procedure leads to the correct determination of $w(z)$, in the sense
that
\begin{equation}
  \langle \hat w \rangle (z) \simeq \frac{\sum_{n=1}^N W_z \bigl(z,
  z^{(n)} \bigr) w_0\bigl( z^{(n)} \bigr)}{\sum_{n=1}^N W_z\bigl(z,
  z^{(n)} \bigr)} \simeq w_0(z) \; .
\end{equation}
Further discussion of Eqs.~(\ref{gamma}) and (\ref{w}) is given in
App.~C.

\subsection{General case}

If the lens is not weak the solution of our problem is more
complicated and requires nesting two different iteration
processes. Let us start by assuming that the cosmological weight
is known. A relation between $\gamma(\vec\theta)$ and the observed
ellipticities now involves the quantity $\kappa(\vec\theta)$, as can be
easily found from Eqs.~(\ref{g}) and (\ref{<epsilon>-s}):
\begin{equation}
  \hat\gamma(\vec\theta) = - \frac{\displaystyle \sum_{n=1}^N W
  \bigl(\vec\theta, \vec\theta^{(n)} \bigr) \left[ \frac{w\bigl(
  z^{(n)} \bigr)}{1 - w\bigl( z^{(n)} \bigr)
  \hat\kappa(\vec\theta^{(n)})} \right] \epsilon^{(n)} }{\displaystyle
  \sum_{n=1}^N W \bigl(\vec\theta, \vec\theta^{(n)} \bigr) \left[
  \frac{w\bigl( z^{(n)} \bigr)}{1 - w\bigl( z^{(n)} \bigr)
  \hat \kappa(\vec\theta^{(n)}) } \right]^2} \; .  \label{gamma-s}
\end{equation}
It is not difficult to recognize that this equation, which generalizes
Eq.~(\ref{gamma}), leads to the optimal estimation of $\gamma_0$. We
note that the combination of weights is again optimized with regard to
the error on $\gamma$. It is clear that Eq.~(\ref{gamma-s}) can be
used to obtain both the shear $\gamma$ and the mass density $\kappa$,
provided that we know the cosmological weight. To this purpose we can
use an initial guess for $\hat\kappa(\vec\theta)$ (e.g.,
$\hat\kappa(\vec\theta) = 0$) in Eq.~(\ref{gamma-s}), obtaining in
this way a first estimation of $\gamma(\vec\theta)$. From this we can
obtain, using standard methods (see Kaiser 1995, Seitz \& Schneider
1996, and Paper~I and II for a discussion), the corresponding mass
density, up to a constant. Then we can go back to Eq.~(\ref{gamma-s})
with the mass density and obtain a new determination of $\hat\gamma$,
and so on. The method outlined here is similar to that described by
Seitz \& Schneider (1997), with the important difference that they
cannot use the weights $w\bigl( z^{(n)} \bigr)$ in Eq.~(\ref{gamma-s})
since the galaxy redshifts are not taken to be known.

For given values of  $\gamma(\vec\theta)$ and $\kappa(\vec\theta)$ we can
estimate $w(z)$ in a manner similar to the weak lensing limit.
In the general case Eq.~(\ref{w}) is replaced by
\begin{eqnarray}
  \hat w(z) = -&\biggl[ \displaystyle \sum_{n=1}^N&W_z \bigl( z,
  z^{(n)} \bigr) \bigl[ 1 - w\bigl( z^{(n)} \bigr) \kappa \bigl(
  \vec\theta^{(n)} \bigr) \bigr]^{-1} \times {} \nonumber\\
  {}&{}\times&\Re \bigl[ \gamma^* \bigl( \vec\theta^{(n)} \bigr)
  \epsilon^{(n)} \bigr] \biggr] \times {} \nonumber\\ {}
  \times&\biggl[ \displaystyle \sum_{n=1}^N&W_z \bigl( z, z^{(n)}
  \bigr) \left| {\gamma\bigl( \vec\theta^{(n)} \bigr) \over 1 -
  w\bigl( z^{(n)} \bigr) \kappa \bigl( \vec\theta^{(n)} \bigr)}
  \right|^2 \biggr]^{-1} \; . 
  \label{w-s}
\end{eqnarray}
As in the weak lensing limit, the joint determination of $\gamma$ and
$w$ is performed by an iterative process between Eq.~(\ref{gamma-s})
and Eq.~(\ref{w-s}). However, we stress once more that in this case
Eq.~(\ref{gamma-s}) requires a separate iteration for the joint
determination of $\gamma$ and $\kappa$, to be performed between
Eq.~(\ref{gamma-s}) and some kernel analysis for mass reconstruction
(see Paper~II).

The proof that Eq.~(\ref{w-s}) leads to an estimation of $w_0(z)$
would be similar to that for Eq.~(\ref{w}).


\section{Invariance properties}

So far, we have assumed that the shear map $\gamma(\vec\theta)$, the
mass distribution $\kappa(\vec\theta)$, and the cosmological weight
$w(z)$ can be determined uniquely. Unfortunately, two invariance
properties affect the lensing equations and leave the reconstruction
process ambiguous.

The first property is the well known ``sheet-invariance'' (Kaiser \&
Squires 1993; Schneider \& Seitz 1995). In the weak lensing limit,
this affects only the dimensionless mass density $\kappa(\vec\theta)$
and has a simple interpretation. If we add a homogeneous ``sheet'' to
the mass density all equations remain unchanged.

The second invariance property, that we call ``global scaling
invariance,'' is a new feature associated with the process of the
joint determination of the lens mass density and of the cosmological
weight.

\subsection{Weak lensing limit}

In this limit the two invariance properties refer to the
transformations (sheet)
\begin{equation}
  \kappa(\vec\theta) \mapsto \kappa(\vec\theta) + C \; , \quad
  \gamma(\vec\theta) \mapsto \gamma(\vec\theta) \; , \quad 
  w(z) \mapsto w(z) \; ,
\end{equation}
and (global scaling)
\begin{equation}
  \kappa(\vec\theta) \mapsto k \kappa(\vec\theta) \; , \quad
  \gamma(\vec\theta) \mapsto k \gamma(\vec\theta) \; , \quad 
  w(z) \mapsto \frac{w(z)}{k} \; .
\end{equation}
We note that, in the weak lensing limit, the two invariance properties
are {\it decoupled}, in the sense that the first involves only the
dimensionless density and the second is met in the determination of
the cosmological weight (see Sect.~3.1) even without considering the
estimation of the dimensionless density.

\subsection{General case}

In the general case, the two invariance properties are {\it
coupled\/}, because all three basic quantities $\kappa$, $\gamma$, and
$w$ have to be determined jointly. In particular, when the lens is not
weak, a new undesired source of uncertainty in the determination of
the cosmological weight is introduced by the sheet invariance itself.

The sheet invariance arises because of the differential relation
between $\gamma(\vec\theta)$ and $\kappa(\vec\theta)$. Let
us recall the method used to obtain the dimensionless density map
$\kappa(\vec\theta)$ from the shear $\gamma(\vec\theta)$. First the
vector field (Kaiser 1995)
\begin{equation}
  \vec u(\vec\theta) = - \left( \matrix{ \gamma_{1,1} + \gamma_{2,2} \cr
  \gamma_{2,1} - \gamma_{1,2} } \right) 
\end{equation}
is calculated from the shear map. As shown by a simple calculation,
this vector field is the gradient of the mass density, i.e.\ $\nabla
\kappa(\vec\theta) = \vec u(\vec\theta)$. Hence, from the shear
$\gamma$, the density $\kappa$ can be determined only up to a
constant. This is the usual sheet invariance in the weak lensing
limit, where the shear is the observed quantity.

In the general case one cannot measure $\gamma(\vec\theta)$ directly
because Eq.~(\ref{gamma-s}), used to obtain the shear, involves the
dimensionless mass density. In addition, a simple generalization to a
relation $\nabla \tilde\kappa = \tilde u$ (see Kaiser 1995 and
Paper~II) is not available because of the presence of the cosmological
weight (see, however, Seitz \& Schneider 1997 for an approximate
relation). Thus an iterative process is needed.  In the iteration, the
arbitrary constant used to invert the vector field $\vec u$ into the
density map $\kappa$ and the cosmological weight act in a complicated
manner, which is hard to describe analytically. If the lens is
sub-critical and if the cosmological weight is left unchanged (i.e.\
if the outer iteration is not performed, see Sect.~3), the sheet
invariance can be expressed by (see Appendix~A)
\begin{equation}
  \kappa \mapsto \frac{C \langle w \rangle_z}{\langle w^2
  \rangle_z} + (1 - C) \kappa(\vec\theta) \; , \quad \gamma(\vec\theta)
  \mapsto (1 - C) \gamma(\vec\theta) \; ,
  \label{fischio/fiasco}
\end{equation}
where $\langle \cdot \rangle_z$ is the average over the redshifts. A
similar expression has been found by Seitz \& Schneider (1997), by
considering the sheet invariance in a different context (the lens
reconstruction for a population of galaxies with known redshift
distribution). Unfortunately, even under the above special conditions,
if one introduces the transformation (\ref{fischio/fiasco}) in
Eq.~(\ref{w-s}), the resulting measured cosmological weight $w(z)$
changes in a complicated manner. It can be shown that the related
transformation for the cosmological weight cannot be reduced to a
simple scaling $w(z) \mapsto \nu w(z)$.

The global scaling invariance property is a the same as that
encountered in the weak lensing limit; but here we recall that the
three quantities $\gamma$, $w$, and $\kappa$ have to be determined
jointly. In particular, if we consider the transformation
\begin{equation}
  \kappa(\vec\theta) \mapsto k \kappa(\vec\theta) \; , \quad
  \gamma(\vec\theta) \mapsto k \gamma(\vec\theta) \; , \quad
  w(z) \mapsto \frac{w(z)}{k} \; ,
\end{equation}
all equations remain unchanged. We observe that a similar invariance
property is met when one derives information on the geometry of the
universe from the separation of multiple images produced by a strong
lens.

\subsection{General strategy}

In conclusion, there are two invariance properties, sheet and
global scaling invariance. In the general case, both invariance
properties affect the determination of the cosmological weight $w(z)$:
the global scaling acts as a simple scaling, while the sheet
invariance acts in a different and a priori unknown manner. Therefore,
to the extent that the impact of the sheet invariance remains out of
control, any determination of the cosmological weight would be
useless. This implies that when the lens is not weak we can choose
between two possibilities. Either we discard the central part of the
cluster, leaving only the parts where the lens is weak, or we must
find a way to break the sheet invariance.

As discussed by Seitz \& Schneider (1997), there are several possible
ways to break the sheet invariance. A simple and useful method is
based on the {\it magnification effect\/} (Broadhurst {\it et al}.\
1995), i.e.\ on the observed (local or total) number of galaxies. In
the case of a single redshift source population, a simple expression
holds (see Appendix~B) for the total projected mass associated with
$\kappa$
\begin{equation}
  M = \int \kappa(\vec\theta) \, \diff^2 \theta = \frac{1}{2} \int
  \bigl[ 1 - \det A(\vec\theta) \bigr] \, \diff^2
  \theta \; .
  \label{m}
\end{equation}
This has a very simple interpretation. Suppose that galaxies are
distributed on the source plane with constant number density
$\rho^\mathrm s$. Then, the observed number density $\rho(\vec\theta)$
(number of galaxies per square arcsec) is given by
\begin{equation}
  \rho(\vec\theta) = \rho^\mathrm s \bigl| \det A(\vec\theta) \bigr| =
  \frac{\rho^\mathrm s}{\mu(\vec\theta)} \; .
\end{equation}
Inserted in Eq.~(\ref{m}) it gives
\begin{eqnarray}
  \hat M &=& \frac{1}{2\rho^\mathrm s} \int \bigl[ \rho^\mathrm s -
  \rho(\vec\theta) \sgn A(\vec\theta) \bigr] \, \diff^2 \theta
  \nonumber\\
  &=& \frac{N_\mathrm{exp} - (N_\mathrm I + N_\mathrm{III} -
  N_\mathrm{II})}{2\rho^\mathrm s} \; .
\end{eqnarray}
In this expression $\sgn A(\vec\theta) = \sgn \bigl(\det
A(\vec\theta)\bigr)$ is the sign of the determinant of
$A(\vec\theta)$, $N_i$ is the observed number of images of kind $i$
(see Schneider {\it et al}.\ 1992 for the relevant classification of
images), and $N_\mathrm{exp}$ the expected number of galaxies in an
area equal to that used for the observations. (We note that the
classification of an observed image is easy once the critical lines
can be determined empirically.)

In view of the above discussion, in the following we will always
consider either the weak lensing limit or the general case with the
sheet mass degeneracy broken. Simulations show that, for the present
problem, a (not too ``strange'') lens with $\kappa \lesssim 0.3$
everywhere can be considered to be weak, i.e.\ to meet the asymptotic
requirements $\kappa, | \gamma | \ll 1$. The global scaling
invariance, instead, will be addressed explicitly in the analysis,
i.e.\ it will be considered present and not broken.

One might think that the global scaling invariance could be broken
simply by the use of the asymptotic behaviour of $w(z)$ at $z
\rightarrow \infty$, i.e.\ by setting the cosmological weight equal to
unity at high redshifts. Unfortunately, the asymptotic regime is
reached only slowly. Thus only galaxies at very high redshifts (say $z
\gtrsim 10$) would be useful, which is practically beyond the limits
of observations. Still, in the discussion of the simulations (see
Sect.~6), we will show that the fact that the global scaling
invariance, in contrast with the sheet invariance, can be reduced to a
simple scaling, allows us to proceed to a meaningful determination of
the cosmological parameters (once the sheet invariance is assumed to
be broken). In fact, we will fit the {\it shape\/} of the measured
cosmological weight with a set of cosmological weights (corresponding
to different cosmological parameters), leaving unspecified the {\it
scale\/} of the function.


\section{Limits on the determination of the cosmological weight function}

In principle, given a determination of the cosmological weight
function $w(z)$ from the procedure outlined in Sect.~3, it is
straightforward to derive information on the parameters $\Omega$ and
$\Omega_\Lambda$ that characterize the assumed Friedmann-Lema\^\i tre
cosmology. In particular, we may resort to a simple least squares or
maximum likelihood method. Such a method is able to lead not only to
``point estimations'' of the cosmological parameters, but also to
``interval estimations,'' based on the identification of the relevant
``confidence level'' regions (see Appendix~D). In practice, difficulties
arise because the determination of $w(z)$ will remain incomplete:
\begin{itemize}

  \item The {\it scale\/} of the cosmological weight is expected to be
  {\it unknown}, because of the global scaling invariance described in
  Sect.~4.

  \item The measurement of $w(z)$ is going to be restricted to a {\it
  limited redshift interval\/} and we anticipate to be able to do so
  with only {\it limited redshift resolution}.

  \item The measurement of $w(z)$, resulting from the study of a
  finite number of source galaxies, is going to be affected by {\it
  statistical errors}.

\end{itemize}
Therefore, we now need to determine, for every cosmological model,
both the expected {\it measured\/} cosmological weight $\langle w
\rangle (z)$ and the related error (covariance matrix). As we
described in Paper~II, since we are dealing with {\it functions},
either of space $\kappa(\vec\theta)$ or of redshift $w(z)$, the
covariance matrix analysis has to be extended to the related two-point
correlation functions. These calculations can be carried out either
analytically (see Appendix~C) or numerically, through Monte Carlo
simulations. In any case, we may anticipate that while an error on the
expected measured $\langle w \rangle (z)$ introduces a {\it bias\/} in
the estimation of the cosmological parameters, an error on the
(co)variance of the cosmological weight is only bound to increase the
expected variance of the cosmological parameters thus determined,
which is not critical for point estimations.


\section{Simulations}

\subsection{Method}

We refer to a cluster at redshift $z_\mathrm d = 0.8$. Different
geometries of the universe set different values for the critical
density $\Sigma_\mathrm c^\infty$. In particular, calling $h = H_0 /
(100 \mbox{ km s}^{-1} \mbox{ Mpc}^{-1})$ the reduced Hubble constant,
the case $\Omega = 1$, $\Omega_\Lambda = 0$ has $\Sigma_\mathrm
c^\infty = 5.49 \, h \mbox{ kg m}^{-2}$, while the case $\Omega =
0.3$, $\Omega_\Lambda = 0.7$ has $\Sigma_\mathrm c^\infty = 3.99\, h
\mbox{ kg m}^{-2}$. In the following we specify the cluster mass
distribution directly in dimensionless form. For this we consider the
sum of a diffuse component A (a truncated Hubble profile) and three
compact objects B, C, D. The corresponding (projected) two-dimensional
distributions are
\begin{equation}
  \kappa = \cases{ K_0 \frac{\displaystyle{\sqrt{1 - r^2 / R^2}}}{
  \displaystyle{1 + r^2 / r_0^2}} & for $r < R \; ,$ \cr
  0 & otherwise$\; ,$ \cr}
\end{equation}
and
\begin{equation}
  \kappa = \frac{K_0}{1/r_0 - 1/R} \left( \frac{1}{\sqrt{r_0^2 + r^2}}
  - \frac{1}{\sqrt{R^2 + r^2}} \right) \; .
\end{equation}
Here $r$ denotes the angular distance from the center of each
object. For simplicity of notation, we use the same symbols $K_0$,
$r_0$, $R$ for three scales, taken to be different in each case (see
Table~1). The dimensionless density $\kappa(\vec\theta)$ is calculated
on a grid of $50 \times 50$ points (corresponding to a field of
$5\arcmin \times 5\arcmin$ on the sky). The cluster, shown in Fig.~2,
has a mass (for $h = 0.5$, $\Omega = 1$, and $\Omega_\Lambda = 0$)
within a $1 \, \mbox{Mpc}$ aperture diameter of about $2 \times
10^{15} \, \mbox{M}_\odot$. From the lens equations we calculate the
shear $\gamma$ on the same grid. Within the Friedmann-Lema\^\i tre
geometry, the weight function is calculated directly from Eqs.~(3),
(4), and (5). For the purpose, $25$ equally spaced points in redshift
from $z_\mathrm d = 0.8$ to $z_\mathrm d = 5.8$ are found to be
sufficient.

\begin{table}[!b]
  \caption{The parameters used for the four components of the cluster
  mass distribution. Component A is a Hubble profile, components B,
  C, and D represent compact objects (``dominant'' galaxies).}
  \small
  \begin{tabular}{c r @{$\arcmin\;$} r @{$\arcsec$} c r @{$\arcmin\;$} 
  r @{$\arcsec$} c r @{$\arcmin\;$} r @{$\arcsec$} c r @{$\arcmin\;$} 
  r @{$\arcsec$} c c}
    \hline
    \hline
     &\multicolumn{6}{c}{Center}&\multicolumn{3}{c}{Inner scale}&%
     \multicolumn{3}{c}{Outer scale}&Density\\
     Comp.& \multicolumn{3}{c}{$\theta_1$} &
     \multicolumn{3}{c}{$\theta_2$} & \multicolumn{3}{c}{$r_0$} &
     \multicolumn{3}{c}{$R$} & $K_0$ \\
     \hline
     A &$ 0$&$ 0$&&$ 0$&$ 0$&&\hskip0.3cm$1$&$15$&&\hskip0.5cm$4$&$0$&&$0.5$\\
     B &$ 0$&$ 0$&&$+0$&$30$&&$0$&$ 9$&&$0$&$30$&&$0.2$\\
     C &$-1$&$30$&&$-1$&$ 0$&&$0$&$15$&&$1$&$ 0$&&$0.3$\\
     D &$+1$&$ 0$&&$-0$&$30$&&$0$&$ 9$&&$0$&$30$&&$0.1$\\
     \hline
     \hline
  \end{tabular}
\end{table}

The ellipticities of source galaxies are generated using a truncated
Gaussian distribution
\begin{equation}
  p \bigl( |\epsilon^\mathrm s| \bigr) = {1 \over \pi P^2 \bigl[ 1 -
  \exp(-1/P^2) \bigr]} \exp \Bigl( - |\epsilon^\mathrm s|^2 / P^2
  \Bigr) \; ,
\end{equation}
with $P = 0.15$. Note that the distribution is normalized with the
condition $2 \pi \int p \bigl( | \epsilon^\mathrm s | \bigr) |
\epsilon^\mathrm s | \, \diff | \epsilon^\mathrm s | = 1$. Galaxy
positions are chosen randomly on the field. For simplicity, following
Seitz \& Schneider (1996), we take a uniform distribution in the {\it
observed\/} position, thus neglecting magnification effects. Finally,
following Brainerd {\it et al}.\ (1996), we draw the redshifts from a
gamma distribution
\begin{equation}
  p_z(z) = {z^2 \over 2 z_0^3} \exp(-z/z_0) \; ,
\end{equation}
where $z_0 = 2/3$. This distribution has maximum at $z = 2 z_0 = 4/3$
and mean $\langle z \rangle = 3 z_0 = 2$. Current surveys favor a
slightly higher value of $z_0$ (Driver {\it et al}.\ 1998, for a
limiting magnitude $I < 26$). All distributions can be obtained from a
uniform distribution by the transformation or rejection methods (see
Press {\it et al}. 1992).


Then the observed galaxy ellipticities are calculated using
Eq.~(\ref{epsilon-s}). The ``observations'' used for the
reconstruction process are then the set of galaxy positions $\bigl\{
\vec\theta^{(n)} \bigr\}$, the set of observed ellipticities $\bigl\{
\epsilon^{(n)} \bigr\}$, and the set of redshifts $\bigl\{ z^{(n)}
\bigr\}$. The simulations are carried out with Gaussian weights in $\|
\vec\theta - \vec\theta' \|$ (with scale lenght $9\arcsec$) and in $|
z - z' |$ (with scale $0.2$).

The reconstruction process follows the method described in Sect.~3,
and can be summarized in 9 steps:

\begin{enumerate}

  \item To begin the reconstruction, an Einstein-de~Sitter universe is
    assumed, and thus, as first guess, the cosmological weight
    $w^{[0]}(z)$ is calculated (Eq.~(\ref{EdS})).

  \item The initial density $\kappa^{[0,0]}$ and the initial shear
    $\gamma^{[0,0]}$ are both set to zero.

  \item A first determination of the shear
    $\gamma^{[0,1]}(\vec\theta)$ is then obtained from the observed
    ellipticities using Eq.~(\ref{gamma-s}).

  \item The new mass distribution $\kappa^{[0,1]}(\vec\theta)$ is
    calculated using the shear $\gamma^{[0,1]}(\vec\theta)$. The
    inversion from the shear into the mass distribution is performed
    by convolution with the curl-free kernel $\vec H^{\rm
    SS}(\vec\theta, \vec\theta')$ (Seitz \& Schneider 1996; for a
    discussion on the merits of different kernels see Paper~II). The
    sheet invariance is broken by requiring the total reconstructed
    mass $\int \kappa(\vec\theta) \, \diff^2 \theta$ to be equal to
    the true value (see Sect.~4.3).

  \item Points (2), (3), and (4) are iterated. This is
    repeated a number $I$ of times sufficient to obtain good
    convergence (typically $I = 5$ for sub-critical clusters).

  \item The final determinations $\kappa^{[0,I]}$ and $\gamma^{[0,I]}$
    are used in Eq.~(\ref{w-s}) to get the new weight function
    $w^{[1]}$.

  \item Points (2--5) are iterated, thus leading to the determination
    of $w^{[j+1]}$ from the knowledge of $w^{[j]}$. (The {\it outer
    iteration\/} on $j$ involves $w$, while the {\it inner
    iteration\/} on $i$ involves $\kappa$ and $\gamma$.)\@ As starting
    density and shear for the new inner iteration we use the last
    determinations of these quantities, i.e.\ $\kappa^{[j+1, 0]} =
    \kappa^{[j, I]}$ and similarly $\gamma^{[j+1, 0]} = \gamma^{[j,
    I]}$. The outer iteration is performed a number $J$ of times.

  \item The final lens density $\kappa^{[J,I]}$ is then compared to
    the true one $\kappa_0$.

  \item The final cosmological weight $w^{[J]}$ is then used to
    determine, or at least constrain, the geometry of the
    universe. The methods used are the ones described in detail in
    Appendix~D. 

\end{enumerate}

Simulations show that, for ``reasonable'' values of $\Omega$ and
$\Omega_\Lambda$, the factor $k$ related to the global scaling
invariance remains very close to unity; thus the measured cosmological
weight does not differ significantly from the true cosmological
weight. This happens because the cosmological weight changes only
slightly for different values of $\Omega$ and $\Omega_\Lambda$ (cfr.\
Figure~1). Hence the initial choice $w^{[0]}(z)$ (Einsten-de~Sitter
universe) is not far from the true $w_0(z)$. In this situation it is
correct to use the measured total mass to break the sheet invariance,
as done in point 4 (instead, in the presence of a wrong scaling factor
$k$, this procedure might lead to undesired results because of the
coupling of the sheet invariance with the global scaling
invariance).

\subsection{Results}

A wide set of simulations show that the method produces significant
results if the number of galaxies studied is a few thousand or more. 

\begin{figure}[!t]
  \resizebox{\hsize}{!}{\includegraphics{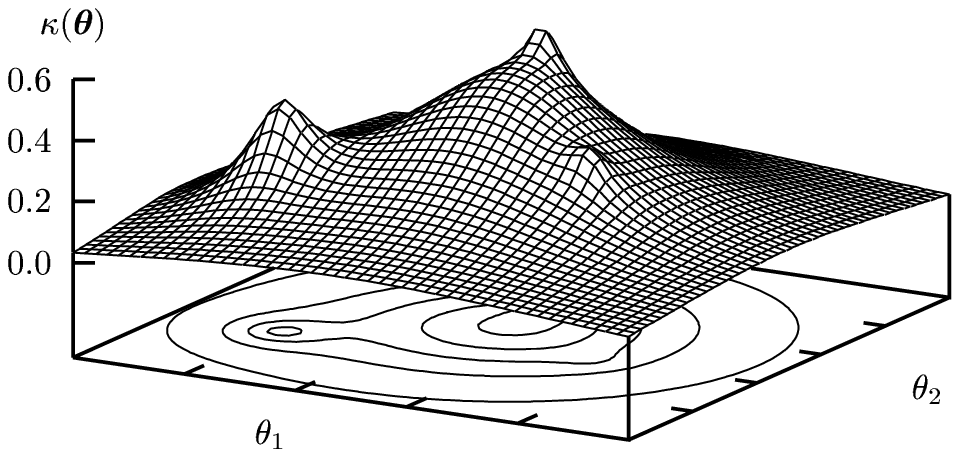}}
  \resizebox{\hsize}{!}{\includegraphics{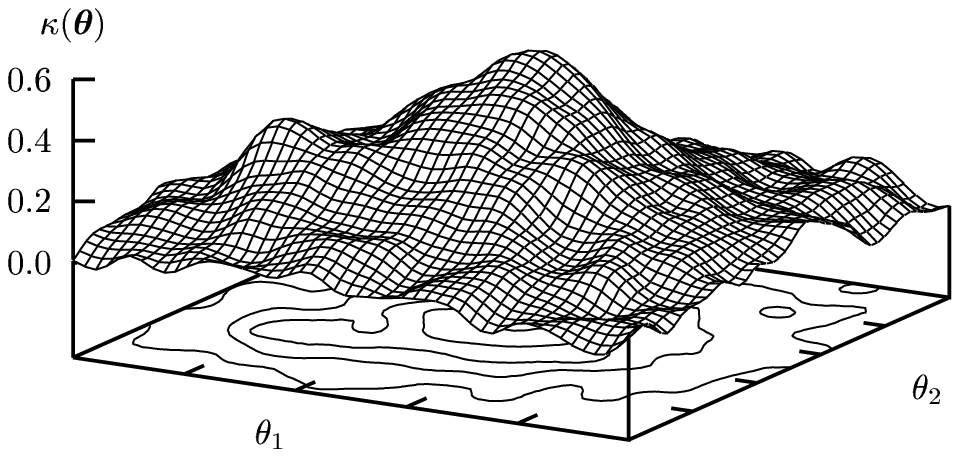}} 
  \caption{The original dimensionless mass density and an example of
    reconstruction. The number of galaxies used is $N=20\,000$; the
    assumed cosmological model is Einstein-de~Sitter. The cluster
    occupies a square with side of $5\arcmin$, corresponding
    approximately to $1.23 \, h^{-1} \mbox{ Mpc}$.}
\end{figure}

The first part of the reconstruction deals with the shear
$\gamma(\vec\theta)$ and with the mass distribution
$\kappa(\vec\theta)$. Figure~2 shows an example of mass density
obtained from the method. The map obtained is rather detailed and
smooth because of the large number of galaxies used ($N=20\,000$).  As
explained in detail in Paper~II, these properties depend on the
spatial weight function $W(\vec\theta, \vec\theta')$ used: the larger
the characteristic size of this function, the smoother the shear and
density maps obtained. This behavior is consistent also with the
calculations of Appendix~C. In the simulations this size has been
adjusted to the value that gives the best results on the following
discussion of the cosmological parameters.

The initial guess for the cosmological weight function $w^{[0]}$ is
not very important for the determination of the shear, because a wrong
choice would only lead to an increase of the errors (at least in the
weak lensing limit; see comment after Eq.~(\ref{K})). This behavior is
well verified in the simulations. At the end of the first inner
iteration, the mass density obtained, $\kappa^{[0,I]}$, does not
differ significantly from the determinations obtained at the end of
the following cycles, $\kappa^{[j, I]}$ (except for a factor arising
from the global scaling invariance). After the first inner cycle, the
mass and the shear maps are close to the best ones that we can hope to
have. This suggests that the number of inner iterations $I$ could
decrease after the first cycle. Thus we could start with $I=5$ for
$j=0$, and then let $I=3$ or even $I=2$ for $j > 0$. The reduction of
$I$ can lead to a significant reduction of machine-time.

\begin{figure}[!t]
  \resizebox{\hsize}{!}{\includegraphics{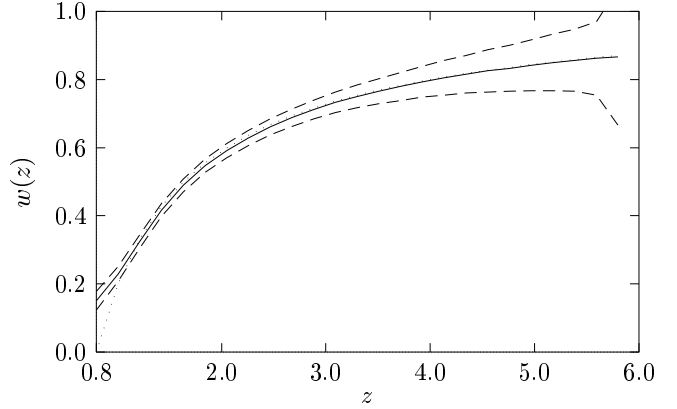}} 
  \caption{The weight expected mean and error on the weight function
  for an Einstein-de~Sitter universe. The solid line shows $\langle w
  \rangle(z)$, the dashed lines the expected errors (obtained as
  $\sqrt{\Cov(w; z, z)}$) and the dotted line the true $w_0(z)$. The
  errors are calculated for $N = 10\, 000$ galaxies; errors for
  different values of $N$ can be obtained by recalling that the error
  scales as $\sqrt{N}$.}
\end{figure}

\begin{figure}[!t]
  \resizebox{\hsize}{!}{\includegraphics{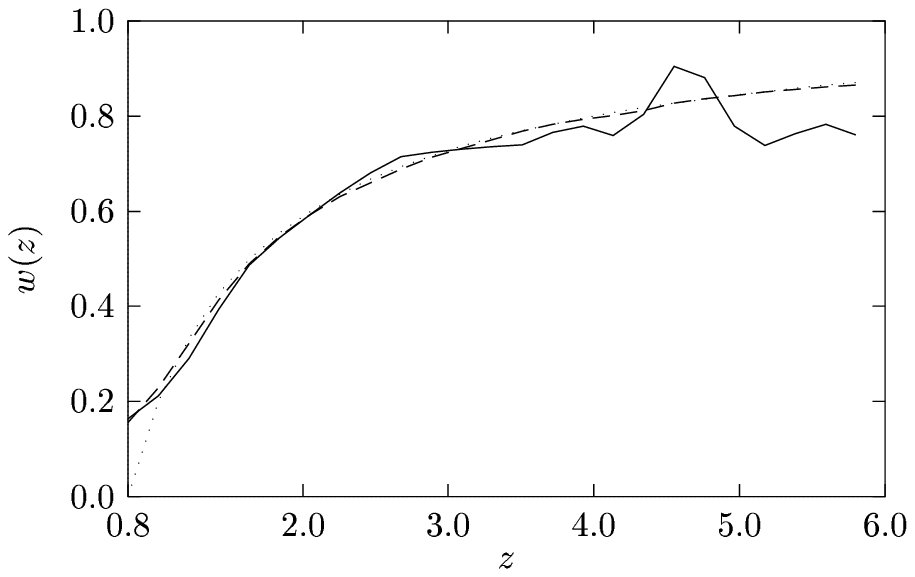}}
  \resizebox{\hsize}{!}{\includegraphics{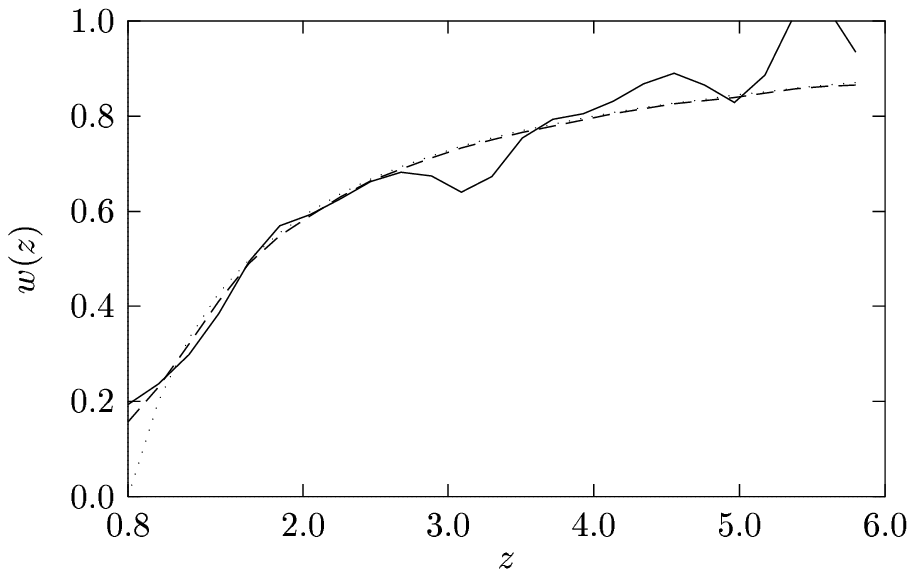}}
  \resizebox{\hsize}{!}{\includegraphics{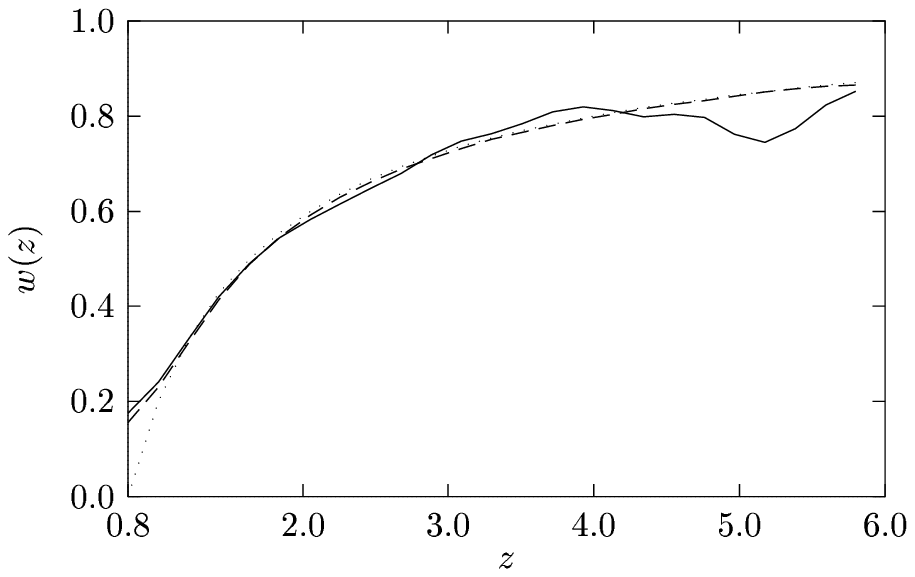}}
  \caption{The reconstructed weight function in an Einstein-de~Sitter
  universe ($\Omega = 1$, $\Omega_\Lambda = 0$). The dotted line
  represent the true weight function $w_0(z)$; the dashed line is the
  expected mean value $\langle w \rangle (z)$, as predicted by
  Eq.~(\ref{<w>}); the solid line is the measured $\hat w(z)$. The
  three figures have been obtained with $N = 10\,000$ (top), $N =
  20\,000$ (middle) and $N = 50\,000$ (bottom) of source galaxies.}
\end{figure}


\begin{figure}[!p]
  \includegraphics{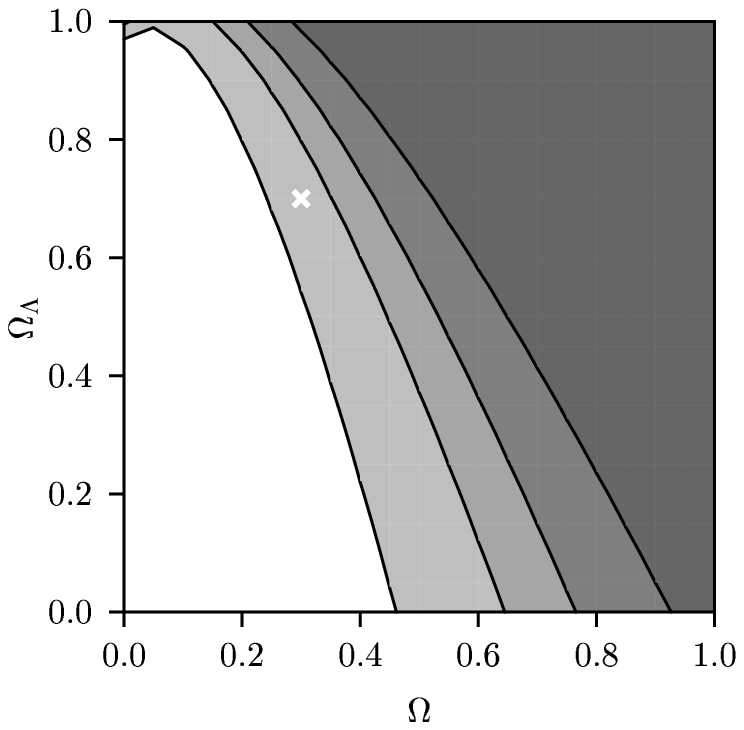}
  \includegraphics{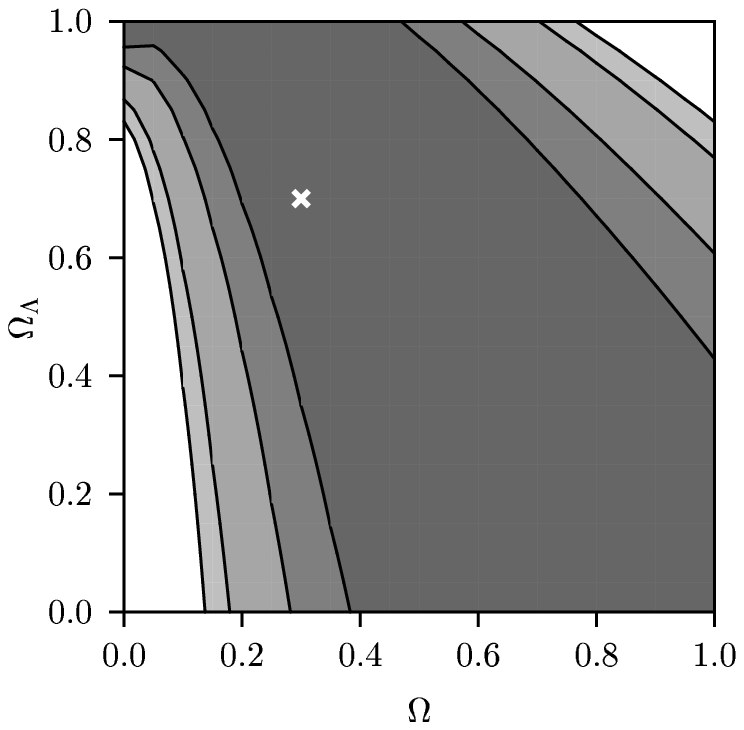}
  \includegraphics{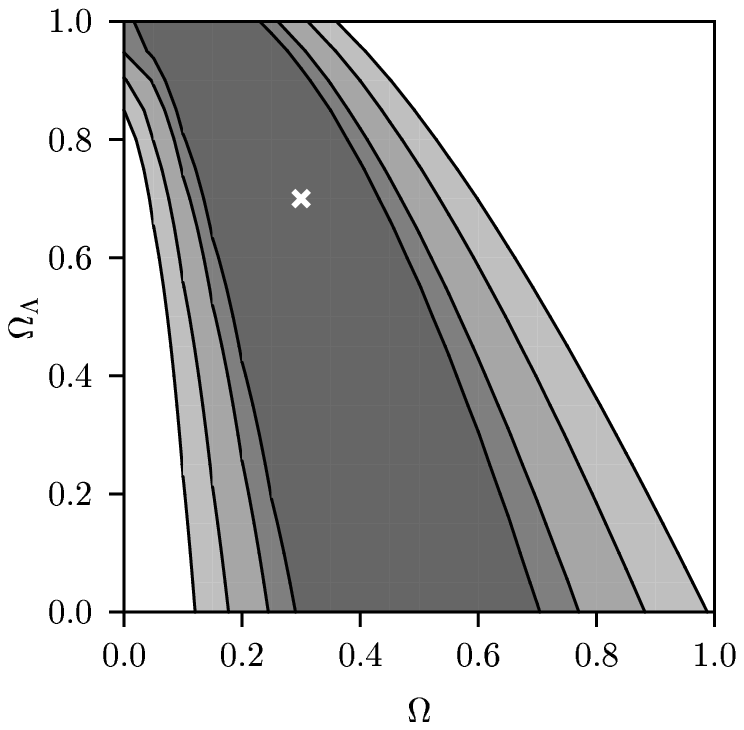}
  \caption{The confidence regions corresponding to $\CL=68\%$,
    $\CL=80\%$, $\CL=90\%$ and $\CL=95\%$. The true value of the
    cosmological parameters, $\Omega = 0.3$ and $\Omega_\Lambda =
    0.7$, is shown as a white cross. The galaxies used are $N = 10\,
    000$ (top), $N = 20\, 000$ (middle) and $N = 50 \, 000$ (bottom).}
\end{figure}

\begin{figure}[!p]
  \includegraphics{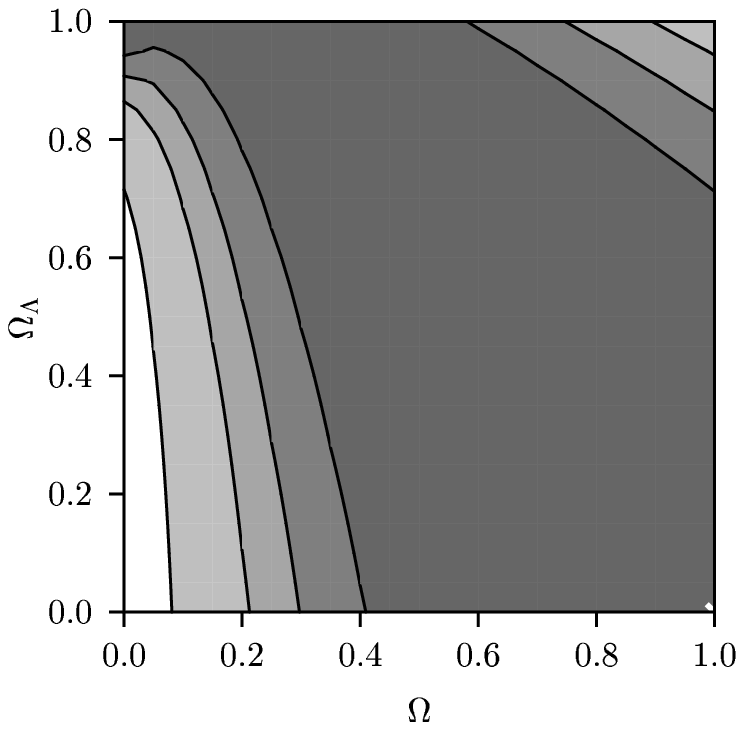}
  \includegraphics{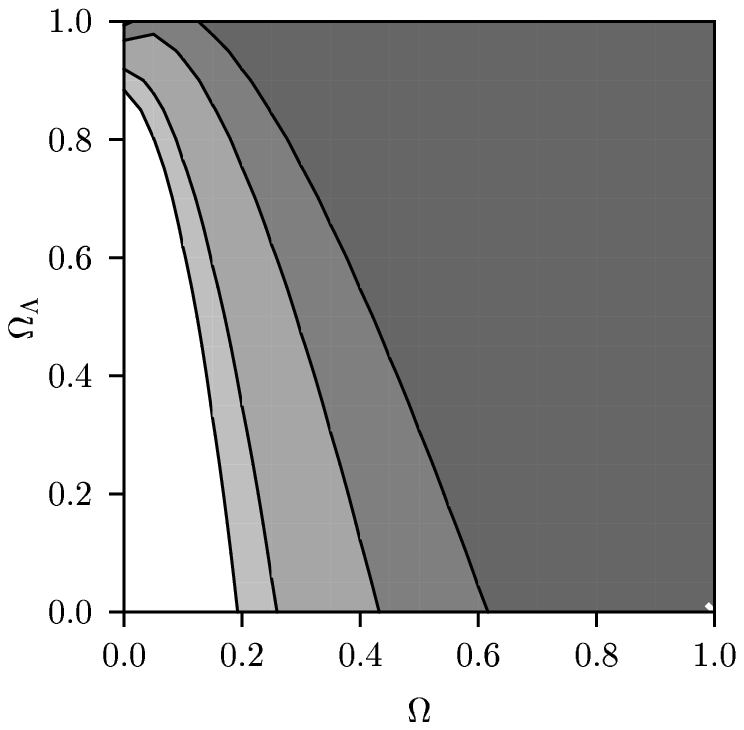}
  \includegraphics{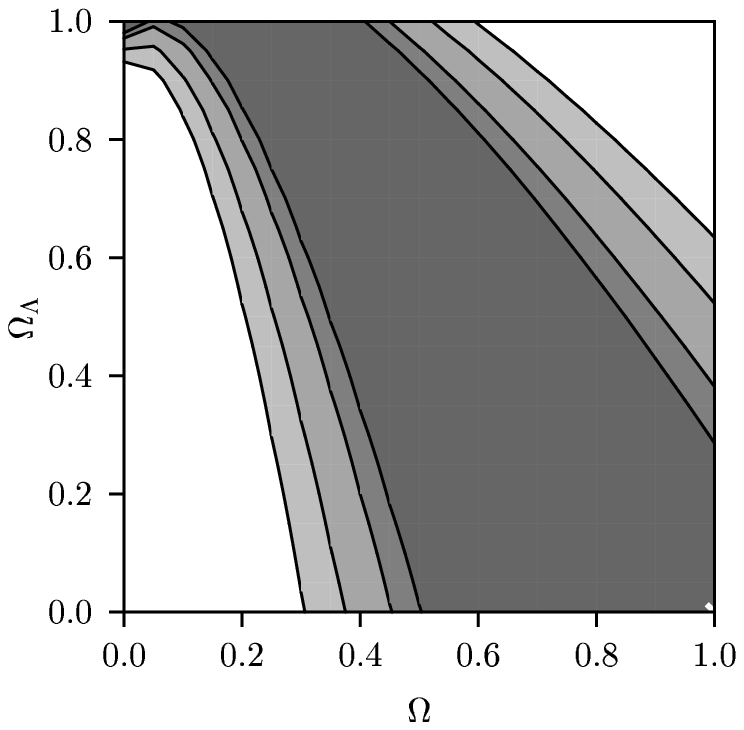}
  \caption{As Fig.~5, but with $\Omega = 1$ and $\Omega_\Lambda =
  0$.}
\end{figure}

The following step is the determination of the cosmological
weight. Simulations show that the number of outer iterations needed to
obtain a good estimation of the cosmological weight is very low, say
$J = 3$, because the method is able to give the correct shear map
after the first inner loop. Simulations also show that this property,
strictly expected only in the weak lensing limit, is in fact valid (at
least approximately) in the general case, provided the lens is not too
``strong.'' The estimated cosmological weight $\hat w(z)$ (solid lines
in Fig.~4) is smooth on the characteristic scale of $W_z(z, z')$ (see
Eq.~(\ref{<w>})); because of this, it remains at a finite value at $z
= z_\mathrm d$. Moreover, as expected, the error on the cosmological
weight increases slightly for $z$ near $z_\mathrm d$ and more for high
values of $z$, where the number of galaxies decreases (see Fig.~3). In
particular, the smoothing of the cosmological weight is very important
near the cluster, at $z \simeq z_\mathrm d$, where the true
cosmological weight vanishes abruptly. This clearly indicates that
neither the limit $w(z) \rightarrow 1$ for $z \rightarrow \infty$ nor
the limit for $z\rightarrow z_\mathrm d^+$ can be used to break the
global scaling invariance. Figure~4 shows the reconstruction of the
cosmological weight in the case of an Einstein-de~Sitter universe;
different choices for $\Omega$ and $\Omega_{\Lambda}$ lead to similar
results. Figure~4 should be compared with Fig.~3 which shows the
expected mean value $\langle w \rangle (z)$ and the related error for
an Einstein-de~Sitter universe with $N = 10\, 000$ galaxies. Figure~3
clarifies the general properties discussed above and, in more detail,
in Appendix~D, for the expected error on $w(z)$ and suggests that our
method should be able to constrain significantly the cosmological
parameters. In fact, the error on $w(z)$ for $z \simeq 2$ is
sufficiently small to distinguish different consmological models even
if the measured weight are affected by the global scaling invariance.

\begin{figure}[!t]
  \resizebox{\hsize}{!}{\includegraphics{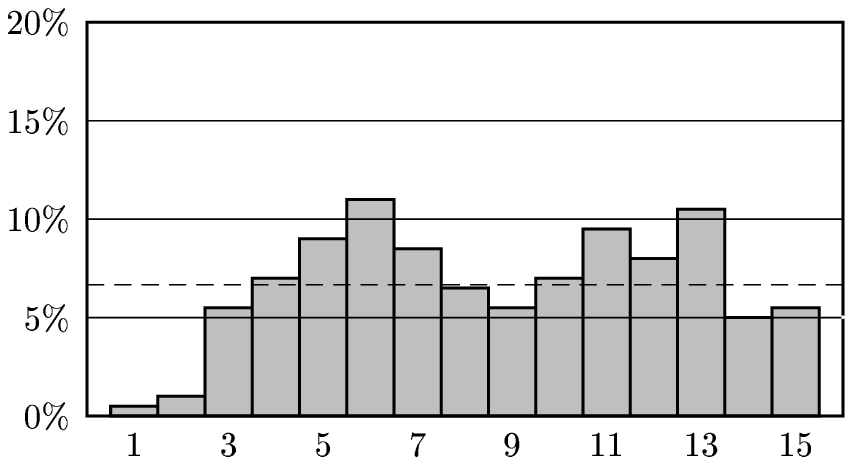}}
  \resizebox{\hsize}{!}{\includegraphics{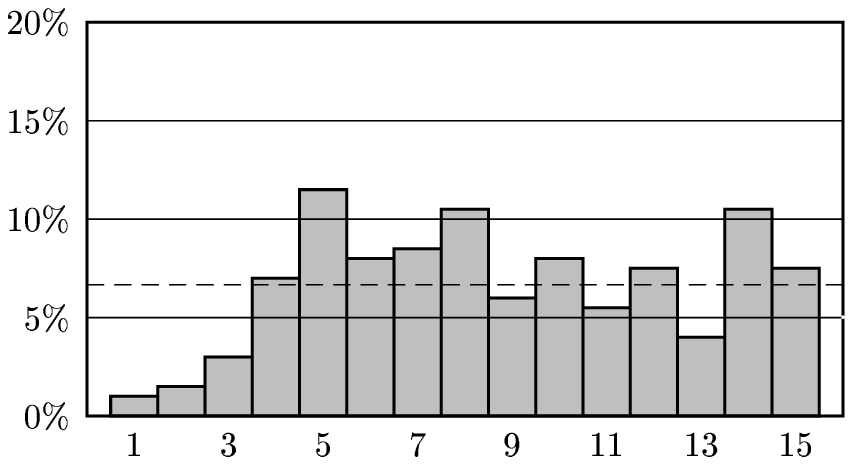}}
  \resizebox{\hsize}{!}{\includegraphics{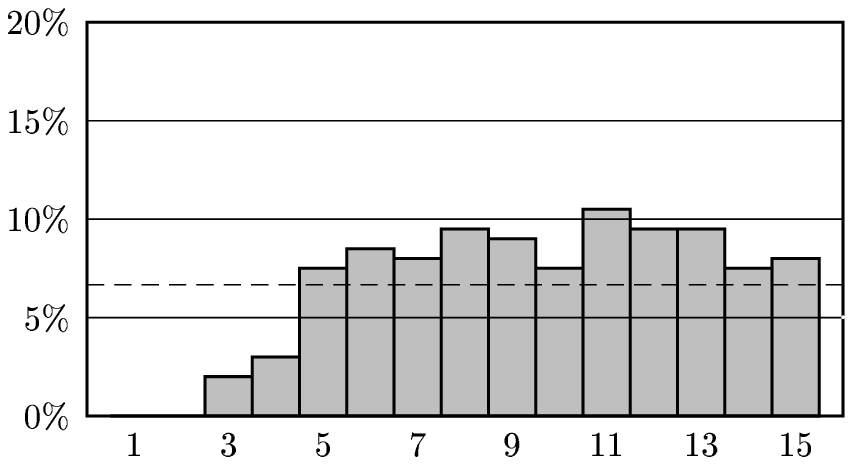}} 
  \caption{The observed probability distribution for $\ell_2$. The 15
  bins have been chosen so that the expected number of events per bin
  would be constant if $\ell_2$ followed a chi-square distribution with
  {\it one\/} degree of freedom. The graphs show that, except for the
  first two bins, the observed probability distribution is roughly
  constant. The three graphs correspond to $N = 10\,000$ (top), $N =
  20\,000$ (middle), and $N = 50\,000$ (bottom).}
\end{figure}

\begin{figure}[!t]
  \resizebox{\hsize}{!}{\includegraphics{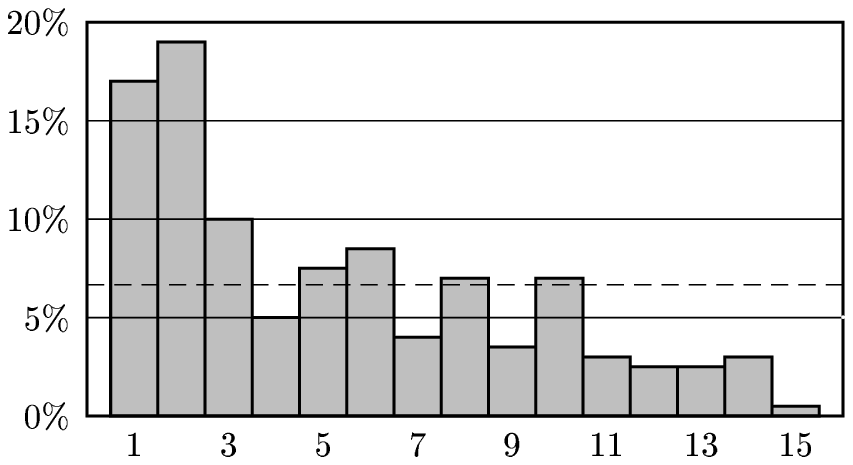}}
  \resizebox{\hsize}{!}{\includegraphics{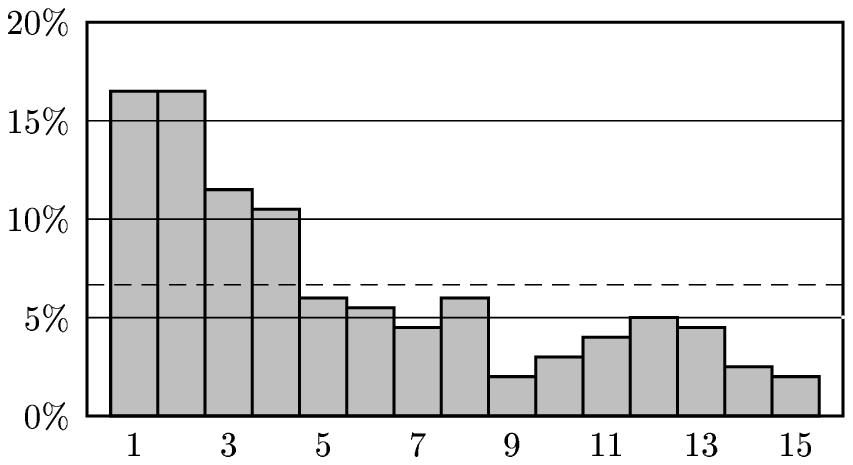}}
  \resizebox{\hsize}{!}{\includegraphics{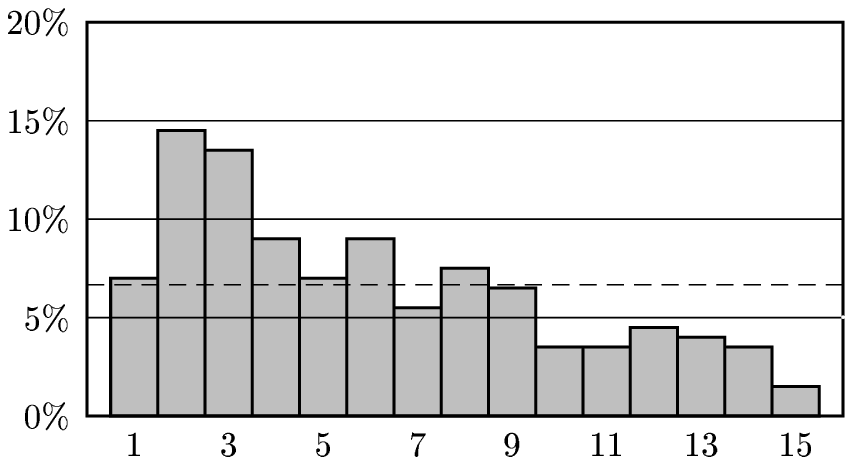}} 
  \caption{As for Fig.~7, but with bins chosen so that the expected
  number of events per bin would be constant if $\ell_2$ followed a
  chi-square distribution with {\it two\/} degrees of freedom.}
\end{figure}

From the estimation $\hat w(z)$ of the cosmological weight we can
obtain information on the cosmological parameters as explained in
Appendix~D. For a description of the results, we plot the contours of the
$\ell_2$ function of Eq.~(\ref{ell2}) in a square domain $[0,1] \times
[0, 1]$ of the $\Omega$-$\Omega_\Lambda$ plane. Figures~5 and 6 show the
confidence regions obtained for various confidence levels $\CL$ in
different cosmological models and with different numbers of source
galaxies. From diagrams of this type we argue that $N = 10\,000$ can
be considered as a lower bound for the applicability of our method.

These figures also show that, unless an exceedingly high number of
source galaxies is available, point estimation is not very
meaningful. In fact, these examples show that the minimum of the
$\chi^2$ function can occur quite far from the true values of the
cosmological parameters. On the other hand, for not unreasonable (see
discussion in Sect.~7) values of $N$, the method does {\it
constrain\/} the cosmological parameters. If additional a priori
conditions are included (e.g., if the universe is taken to have
$\Omega_\Lambda = 0$), parameters can be constrained rather
efficiently.

In order to check the distribution of $\ell_2$, we have repeated a
large number of times the entire process, using each time a different
set of source galaxies. The measured probability distribution of
$\ell_2$ for different values of $N$ is shown in Figs.~7 and 8. In
these figures, the histograms show the number of simulations that have
produced a value of $\ell_2$ in the corresponding interval. Intervals
are chosen so that the theoretical number of events per bin is
constant. From these and other results not shown it is clear that
$\ell_2$ tends to a chi-square distribution with two degrees of
freedom only for a very large number of galaxies $N$, say $N \gtrsim
100\,000$. In contrast, for ``small'' values of $N$ the observed
distribution is (approximately) consistent with a chi-square with one
degree of freedom. This suggests that, when the number of galaxies is
small, a new approximate invariance may be present. The method is able
to break this invariance for larger values of $N$.

This behavior is also suggested by the shapes of the confidence
regions of Figs.~5 and 6. In fact, the confidence regions obtained
have a characteristic shape approximately following the lines $\Omega
+ \Omega_\Lambda = 1 - \Omega_R = \mbox{const}$. Thus our method
should be able to constrain well the curvature $\Omega_R$, while the
quantity $\Omega - \Omega_\Lambda$ should remain practically unknown,
at least when the number $N$ of galaxies used is not too high. This
approximate invariance may explain the behavior of the quantity
$\ell_2$ for finite $N$. In fact, if the invariance were exact,
$\ell_2$ would be expected to follow a chi-square distribution with
(only) one degree of freedom. This is indeed observed for $N \lesssim
50\,000$. Thus in order to draw the confidence regions, one should use
the results of Monte Carlo simulations, unless the number $N$ of
galaxies is {\it very\/} large. The contours presented in this paper
have been obtained from Monte Carlo simulations. The use of the
asymptotic limit would have led indeed to {\it larger\/} confidence
regions.


\section{Feasibility}

In this section we briefly discuss the applicability of our method to
observations. As noted earlier, the main difficulty with our method is
probably the high number of galaxies (and redshifts) needed to obtain
significant constraints on the geometry of the universe.

In order to show that the examples chosen for our simulations can be
considered to be not unrealistic, we first note that the required
density of galaxies is between $400$ and $800$ galaxies per square
arcminute. This density is certainly too high for current
instruments. In fact, the current achievable density of galaxies (with
``reasonable'' integration times) is of the order of $60$ galaxies per
square arcminute (Clowe {\it et al}. 1998), and this number is only
likely to double with the Advanced Camera on HST. However, current
estimates for NGST (Stiavelli {\it et al}. 1997) give more than
$1\,000$ galaxies per square arcminute. It is also interesting to note
that the baseline for the camera currently considered for NGST is
about $4\arcmin$, not far from the value used in our simulations
($5\arcmin$).

A second difficulty with the method proposed here is the large number
of redshifts required. As suggested in the Introduction, photometric
redshifts might be used. In this case, however, sizable statistical
errors on the measured redshifts are expected. At this stage, for
simplicity, such errors have not been taken into account in the
simulations, but they obviously should be considered in detail before
applications are tried. (We note also that errors on the measurement
of ellipticities may have an impact, especially if small and weak
source galaxies are used.)

Another important ingredient is the cluster chosen in our
simulations. In an Einstein-de~Sitter universe with $h = 0.5$, the
cluster would have a mass within a $1 \, \mbox{Mpc}$ aperture of about
$2 \times 10^{15} \, \mbox{M}_\odot$. We may compare our cluster with
the X-ray cluster MS $1054-03$ located at $z_d = 0.83$ (Luppino \&
Kaiser 1997); weak lensing estimates for this cluster give a mass
within $1 \, \mbox{Mpc}$ in the range $1.2 \times 10^{15} \,
\mbox{M}_\odot$ and $5.5 \times 10^{15} \, \mbox{M}_\odot$, depending
on the assumed redshift distribution for the source galaxies. 

In conclusion, the proposed method should give interesting results if
applied to NGST observations. Of course, several refinements and
improvements would be required for the purpose. For example, as
suggested in the Introduction, the use of several clusters should
provide tighter constraints on the cosmological parameters. Broadly
speaking, the use of $N_\mathrm{cl}$ similar clusters should make the
errors on the cosmological parameters smaller by a factor
$\sqrt{N_\mathrm{cl}}$. However, for this purpose the method should be
generalized to be applied to different clusters at different redshifts
(actually, we expect that if the redshifts of the clusters are
different, more information on the geometry of the universe is
available). We should also keep in mind that, intuitively, the use of
several clusters may introduce additional difficulties, much like the
case of a single cluster with pronounced substructure.

Finally, we should stress that a potential problem for the application
of our method could be that of multiple lensing. In fact, if weak
lensing analysis is used (in order to solve the problem of the sheet
invariance), ``tidal'' effects of other clusters (or from large scale
structures) could become important and may jeopardize the
applicability of the method. On the other hand, it is often very hard
to break the sheet invariance. In view of the above considerations, it
would be very interesting to develop a method able to overcome these
problems, possibly by decoupling the sheet and the global scaling
invariance properties.


\section{Conclusions}

In this paper we have proposed and discussed a method, based on weak
lensing and redshift observations, to reconstruct the lens mass
distribution and to obtain, at the same time, information on the
geometry of the universe. Simulations based on the homogeneous
Friedmann-Lema\^\i tre models have shown that the method is viable
when the number of source galaxies is sufficiently high. In principle,
the same method could be extended to other cosmological models,
provided that the angular diameter-distance relation be available;
some of these would be harder to handle because they involve a larger
number of free parameters (e.g., the inhomogeneous ``in average''
Friedmann-Lema\^\i tre models; see Kayser {\it et al}.\ 1997). 

In this paper we have used a ``parameter-free'' approach in the
determination of the cosmological weight. In fact, in the
reconstruction process we have tried to determine the values of $w(z)$
for a number of different redshifts (25 in our simulations), without
any assumption on its shape. A different method could have been used,
namely the direct determination of the cosmological parameters without
the intermediate determination of the cosmological weight $w(z)$
(parametric approach). In this case the cosmological parameters could
have been estimated using, for example, a maximum likelihood method. A
similar dichotomy exists between parametric and non-parametric mass
reconstructions. We have decided to use here the non-parametric
approach as more ``conservative,'' but it would be interesting to
develop a parametric method in detail for comparison.

One improvement that could be made on the current method is based on
more direct use of the shear in the determination of the cosmological
weight (see Eq.~(\ref{w})). In fact, the complex shear
$\gamma(\vec\theta)$ is not an independent quantity, in the sense that
a relation between its partial derivatives holds (i.e., $\nabla \wedge
\vec u(\vec\theta) = 0$; see Paper~II for a detailed
discussion). Hence it is natural to expect that the inclusion of an
additional relation, not considered in the present paper, could
improve the determination of $w(z)$. This might be implemented by
resorting to a maximum likelihood approach for the measurement of the
shear map.

\begin{acknowledgements}
  It is a pleasure to thank Peter Schneider for several suggestions
  that have helped us improve the paper.
  This work has been partially supported by MURST and by ASI of
  Italy. The simulations have been performed on a Digital Alpha LX164
  running at 533 MHz.
\end{acknowledgements}

\begin{appendix}

\section{The sheet invariance}

In this Appendix we describe the sheet invariance for non-weak
lenses. Our discussion largely follows Seitz \&
Schneider (1997).

Following the approach of Appendix~C, we start by taking a fixed
cosmological weight equal to the true cosmological weight
times a given constant, i.e.\ $w(z) = \nu w_0(z)$. By using a global
scaling invariance with scaling factor $k = 1 / \nu$ we can make
$w(z) = w_0(z)$. An approximate expression for the mean value of the
measured shear, valid also for non-weak lenses, can be shown to be
\begin{equation}
  \langle \gamma \rangle \simeq \frac{\langle w \rangle_z - \kappa
  \langle w^2 \rangle_z}{\langle w \rangle_z - \kappa_0 \langle w^2
  \rangle_z} \gamma_0 \; .
  \label{A1}
\end{equation}
In this equation we have neglected the smoothing effect of the weight
function $W(\vec\theta, \vec\theta')$. From this relation we can
derive the form of the sheet invariance. Suppose that $\langle \gamma
\rangle = (1 - C) \gamma_0$, with $C$ constant. This simply means that
the measured value of $\gamma$ is proportional to the true shear. Then
the previous equation gives
\begin{equation}
  \kappa = \frac{C \langle w \rangle_z}{\langle w^2 \rangle_z} + (1 -
  C) \kappa_0 \; .
  \label{A2}
\end{equation}
This equation is consistent with the assumption $\gamma = (1 - C)
\gamma_0$ because the ``sheet constant'' $C \langle w \rangle_z /
\langle w^2 \rangle_z$ does not affect the
shear. Equation~(\ref{fischio/fiasco}) thus gives the form of the
sheet invariance for the dimensionless projected density $\kappa$ in
the general case.

If we use these relations in Eq.~(\ref{w-s}) we can derive the
expected transformation induced on the measurement of the cosmological
weight. The expressions obtained, rather complicated, are not reported
here. However it is clear that the cosmological weight {\it
changes\/}, i.e.\ the measured $w(z)$ is different from
$w_0(z)$. Moreover, it can be shown that $w(z) / w_0(z)$ is not
constant: this of course means that the sheet invariance has a
different effect on the cosmological weight with respect to the global
scaling invariance.


\section{A constraint on the total projected mass of the lens}

The goal of this Appendix is to prove Eq.~(\ref{m}) (following
Lombardi 1996). As a first step we note that
\begin{eqnarray}
  \int \bigl[ 1 - \det A(\vec\theta) \bigr] \, \diff^2 \theta &=&
  \lim_{R \rightarrow \infty} \sint{B_R}{} \; \bigl[ 1 - \det A(\vec\theta)
  \bigr] \, \diff^2 \theta \nonumber \\ 
  &=& \lim_{R \rightarrow \infty} \biggl( \sint{B_R}{} \diff^2 \theta -
  \sint{\vec\theta^\mathrm s(B_R)}{} \; \diff^2\theta^\mathrm s
  \biggr) \; .  \label{AppB}
\end{eqnarray}
In this equation $B_R$ is the disk of radius $R$ centered in the
origin of the $\vec\theta$-plane, while $\vec\theta^\mathrm s(B_R)$ is
the image, through the ray-tracing function $\vec\theta^\mathrm
s(\vec\theta)$, of the disk $B_R$. The second equality holds because
of the change of variable formula.

A well-known property of gravitational lenses is that the deflection
$\vec\beta(\vec\theta) = \vec\theta^\mathrm s(\vec\theta) -
\vec\theta$ can be expressed as the gradient of the deflection
potential $\psi(\vec\theta)$ (given the symmetry of the Jacobian
matrix $A$ of Eq.~(\ref{A})). Such potential can be written as
\begin{equation}
  \psi(\vec\theta) = {1 \over \pi} \int \kappa(\vec\theta') \ln \|
  \vec\theta - \vec\theta' \| \, \diff^2 \theta' \; .
\end{equation}
For an isolated distribution of matter, such that the mass density
$\kappa$ vanishes outside a given disk $B_R$, the potential can be
expressed (outside $B_R$) in terms of a multipole expansion 
\begin{equation}
  \psi(\vec\theta) = {1 \over \pi} \biggl( M \ln \theta - {\vec p \cdot
  \vec\theta \over \theta^2} - {1 \over 2} \sum_{ij} {q_{ij} \theta_i
  \theta_j \over \theta^4} - \cdots \biggr) \; ,
\end{equation}
where
\begin{eqnarray}
  M & = & \int \kappa(\vec\theta') \, \diff^2 \theta' \; , \\
  p_i & = & \int \theta'_i \kappa(\vec\theta') \, \diff^2 \theta' \; , \\
  q_{ij} & = & \int (2 \theta'_i \theta'_j - \delta_{ij} \theta'^2)
  \kappa(\vec\theta') \, \diff^2 \theta' \; .
\end{eqnarray}
(The first physically important term after $M$ is the quadrupole
$q_{ij}$, as the dipole $p_i$ vanishes if the origin of axes is chosen
in a suitable way.) Thus we obtain for the deflection
\begin{equation}
  \vec\beta(\vec\theta) = \nabla \psi(\vec\theta) = {1 \over \pi}
  \left( {M \vec\theta \over \theta^2} - {\vec p \over \theta^2} + {2
  (\vec p \cdot \vec\theta) \vec\theta \over \theta^4} +
  \cdots \right) \; .
\end{equation}
Using this expression we can estimate the last integral in
Eq.~(\ref{AppB}). In fact, using the definition of
$\vec\beta(\vec\theta)$ and neglecting terms of order
$\O\bigl(\theta^{-2}\bigr)$, we can say that the region
$\vec\theta^\mathrm s(B_R)$ is a disk of radius
\begin{equation}
  R^\mathrm s = R - {M \over \pi R} \; .
\end{equation}
Thus we find
\begin{eqnarray}
  \sint{\vec\theta^\mathrm s(B_R)}{} \; \diff^2 \theta &=& \pi \left( R
  - {M \over \pi R} \right)^2 + \O\bigl(R^{-1}\bigr) \nonumber \\
  &=& \pi R^2 - 2 M + \O\bigl(R^{-1}\bigr) \; .
\end{eqnarray}
Inserting this expression in Eq.~(\ref{AppB}), we obtain
Eq.~(\ref{m}). A more detailed discussion shows that, in reality, the
error involved is not $\O\bigl(R^{-1}\bigr)$ but
$\O\bigl(R^{-2}\bigr)$, and precisely it is of order $\O\bigl(M^2/R^2,
q/R^2\bigr)$. This property has to do with the fact that the dipole
$\vec p$, which would be responsible for the error of order
$\O\bigl(R^{-1}\bigr)$, can be made to vanish with a suitable choice
of the origin of the axes. However, the origin of coordinates is not
important when we take the limit $R \rightarrow \infty$, and thus the
error is only of order $\O\bigl(R^{-2}\bigr)$.

An improved version of Eq.~(\ref{m}) can be obtained by retaining the
term of order $\O\bigl(M^2/R^2 \bigr)$:
\begin{equation}
  {1 \over 2} \int_{B_R} \bigl[ 1 - \det A(\vec\theta) \bigr] \,
  \diff^2 \theta \simeq M - {M^2 \over 2 \pi R^2} \; .
\end{equation}
This expression may become useful when the quadrupole $q_{ij}$ is
neglegible.


\section{Expectation values and errors}

An analytic calculation of the expectation values and errors for the
shear and for the cosmological weight is important both for a first
check of the quality of our method and for its practical
implementation. The problem is not easy to solve in the general case,
mainly because it involves iterations of different equations. For this
reason we will often simplify the discussion by introducing some
approximations.

We expect calculations for the covariance matrix of $\hat w$ to be
especially difficult, mainly because in Eq.~(\ref{w}) for $\hat w(z)$
the quantity $\gamma(\vec\theta)$ has been calculated using the
{\it same\/} galaxies to be used for $w(z)$. Thus the errors on $\hat
w$ and on $\hat\gamma$ should be correlated. (Strictly speaking this
is true also for the covariance matrix of $\hat\gamma$, but such
dependence is not very important for our problem.) For simplicity, in
the following we will consider either the cosmological weight or the
shear map fixed in the calculation of the expectation value and
error of the other quantity. The results obtained, checked by
simulations, show that this approximation is good.

The general method used for calculating expectation values and errors
is described in detail in Paper~I and Paper~II (see especially
Appendix~C of the latter article) and is based on the assumption that
we can linearize the relevant equations near the mean value of the
random variables. Calculations will be done by replacing summations
with integrals, so that Poisson noise will be neglected. As explained
in Paper~II, Poisson noise generally adds only a small correction to
the dominant sources of error. For this purpose we define the redshift
probability distribution $p_z(z)$. Hence the probability that a galaxy
is observed in the solid angle $\diff^2 \theta$ and with redshift in
the range $[z, z + \diff z]$ is given by $\rho p_z(z) \, \diff^2
\theta \, \diff z$ (notice that source galaxies are taken to be
uniformly distributed on the field, and thus $\rho$ does not depend on
$\vec\theta$). In the following, for simplicity, we will also assume
that the distribution of source ellipticities is independent of
redshift.

\subsection{Weak lensing}

\subsubsection{Expectation values}

Calculations are not difficult in the weak lensing limit because all
expressions are linear. The shear is calculated from
Eq.~(\ref{gamma}), and thus, from Eq.~(\ref{<epsilon>}), the expected
value of the shear is
\begin{equation}
  \langle \hat\gamma \rangle(\vec\theta) = \frac{\sum_{n=1}^N W\bigl(
  \vec\theta, \vec\theta^{(n)} \bigr) w\bigl( z^{(n)} \bigr) w_0
  \bigl( z^{(n)} \bigr) \gamma_0 \bigl( \vec\theta^{(n)}
  \bigr)}{\sum_{n=1}^N W\bigl( \vec\theta, \vec\theta^{(n)} \bigr)
  \bigl[ w\bigl( z^{(n)} \bigr) \bigr]^2 } \; .
\end{equation}
If we move to a continuous distribution we find
\begin{eqnarray}
  \langle \hat\gamma \rangle(\vec\theta) &=& \biggl[ \displaystyle
  \int \!  \diff^2\theta' \! \int \! \diff z' \, p_z(z') W(\vec\theta,
  \vec\theta') w(z') w_0(z') \times {} \nonumber\\ & & {} \times
  \gamma_0(\vec\theta') \biggr] \biggl[ \int \diff^2\theta' \int \diff
  z' \, W(\vec\theta, \vec\theta') \times {} \nonumber\\ & & p_z(z')
  \bigl[ w(z') \bigr]^2 \biggr]^{-1} \; .
\end{eqnarray}
We observe now that the integrals on $\diff^2\theta'$ and on $\diff
z'$ factorize, thus leaving simply
\begin{equation}
  \langle \hat\gamma \rangle(\vec\theta) = K \frac{ \int
  \diff^2\theta' \, W(\vec\theta, \vec\theta') \gamma_0(\vec\theta')
  }{ \int \diff^2\theta' \, W(\vec\theta, \vec\theta') } \; .
  \label{<gamma>}
\end{equation}
The constant $K$ is obviously related to the global scaling invariance
and is given by
\begin{equation}
  K = \frac{ \int \diff z' \, p_z(z') w(z') w_0(z') }{ \int \diff z'
  \, p_z(z') \bigl[ w(z') \bigr]^2 } \; .
  \label{K}
\end{equation}
Note that $K$ is $\vec\theta$-independent, so that the use of an
inaccurate cosmological weight affects only the scale of the expected
measured shear. A result similar to Eq.~(\ref{<gamma>}) has been
obtained in Paper~II.

From Eq.~(\ref{w}), the expectation value of the cosmological weight
is given by
\begin{equation}
  \langle \hat w \rangle (z) = \frac{\sum_{n=1}^N W_z \bigl(z, z^{(n)}
  \bigr) \Re \bigl[ w_0 \bigl( z^{(n)} \bigr) \hat\gamma\bigl(
  \vec\theta^{(n)} \bigr) \gamma_0^*\bigl( \vec\theta^{(n)} \big)
  \bigr]}{ \sum_{n=1}^N W_z\bigl(z, z^{(n)} \bigr) \bigl| \hat\gamma
  \bigl( \vec\theta^{(n)} \bigr) \bigr|^2} \; .
\end{equation}
By moving to a continuous distribution and by replacing $\hat\gamma$
with its expected value given by Eq.~(\ref{<gamma>}), we obtain
\begin{eqnarray}
  \langle \hat w \rangle (z) &=& \frac{1}{K} \biggl[ \int \! \diff z'
  \! \int \diff^2 \theta' \, p_z(z') W_z(z, z') w_0(z') \times
  \nonumber\\ & & \Re \biggl( \int \! \diff^2 \theta'' \,
  W(\vec\theta', \vec\theta'') \gamma_0^*(\vec\theta')
  \gamma_0(\vec\theta'') \biggr) \biggr] \times {} \nonumber\\ 
  & & {} \times \biggl[ \int \! \diff z' \! \int \! \diff^2 \theta' \,
  p_z(z') W_z(z, z') \nonumber\\ 
  & & \left| \int \!  \diff^2 \theta'' \,
  W(\vec\theta', \vec\theta'') \gamma_0(\vec\theta'') \right|^2
  \biggr]^{-1} \; .
\end{eqnarray}
As for the shear the integrals on the redshift factorize. Let us
consider in more detail the integrals on the angles, i.e.\
$\Re(\cdots)$ in the numerator and $\int \diff^2 \theta' \int \diff^2
\theta'' | \cdots |^2$ in the denominator. These two quantities differ
only by one integration with weight $W(\vec\theta', \vec\theta'')$.
%
%
Thus we may argue that
\begin{equation}
  \langle \hat w \rangle (z) \simeq \frac{1}{K} \frac{ \int \diff z' \,
  p_z(z') W_z(z, z') w_0(z') }{ \int \diff z' \, p_z(z') W_z(z, z') }
  \; .  \label{<w>}
\end{equation}
We note that here the constant $K$ appears at the denominator,
consistent with the global scaling invariance.

In the following, in order to simplify equations,
we will introduce the normalization condition
\begin{equation}
  \int \diff z' \, p_z(z') W_z(z, z') = 1 \quad \forall z \; .
  \label{norm2}
\end{equation}
Similarly, we will consider the angular weight functions (see
Eq.~(\ref{<gamma>})) to be normalized, i.e.\ to satisfy the condition
\begin{equation}
  \rho \int W(\vec\theta, \vec\theta') \, \diff^2\theta' = 1 \quad
  \forall \vec\theta \; .  \label{norm1}
\end{equation}
We recall that this does not limit our discussion, since only {\it
relative\/} values of $W$ are important (see Paper~II).

\subsubsection{Errors}

Let us now consider the error (i.e.\ the covariance matrix) of
$\hat\gamma$. Using the definitions of Paper~I and Paper~II and
calling $\epsilon_1$ and $\epsilon_2$ the real and imaginary
components of the ellipticity $\epsilon$, we find
\begin{equation}
  \Cov (\hat\gamma; \vec\theta, \vec\theta') = \sum_{n=1}^N
  \left( \frac{\partial \hat\gamma(\vec\theta)}{\partial
  \epsilon^{(n)}} \right)
  \Cov \bigl( \epsilon^{(n)} \bigr) \left( \frac{\partial
  \hat\gamma(\vec\theta')}{\partial \epsilon^{(n)}} \right)^\mathrm
  T\; .
\end{equation}
In this equation we have used the matrix notation for the partial
derivatives. For example, the last matrix, because of the transpose,
has $\partial \hat \gamma_j(\vec\theta') / \partial \epsilon_i^{(n)}$
in the element with row $i$ and column $j$. From Eq.~(\ref{epsilon})
we find that $\Cov(\epsilon) = \Cov(\epsilon^\mathrm s) = c \,
\mathrm{Id}$, where $c$ is a positive constant and $\mathrm{Id}$ is
the $2 \times 2$ identity matrix (note that, as the source
ellipticities are taken to be independent of redshift, $c$ is
independent of redshift as well). The partial derivatives of
$\hat\gamma$ can be calculated from Eq.~(\ref{gamma}). By moving to
the continuous description, after some manipulations we find
\begin{equation}
  \Cov (\hat\gamma; \vec\theta, \vec\theta') = c \rho \frac{ \int
  \diff^2 \theta'' \, W(\vec\theta, \vec\theta'') W(\vec\theta',
  \vec\theta'') }{ \int \diff z'' \, p_z(z'') \bigl[ w(z'') \bigr]^2 }
  \, \mathrm{Id} \; , \label{Cov(gamma)}
\end{equation}
where the normalization condition (\ref{norm1}) has been used. This
result is a generalization of a similar result obtained in Paper~II
(see Eq.~(26) of that article). 

If we consider $\gamma$ fixed, the covariance matrix of $\hat w$ is
given by
\begin{equation}
  \Cov(\hat w; z, z') = \sum_{n=1}^N \left( \frac{\partial \hat w(z) }{
  \partial \epsilon^{(n)}} \right) \Cov_{ij}\bigl( \epsilon^{(n)} \bigr)
  \left( \frac{\partial \hat w(z') }{ \partial \epsilon^{(n)}}
  \right)^\mathrm T \; .
\end{equation}
The partial derivatives can be calculated from Eq.~(\ref{w}).
Turning to a continuous description and using the normalization
condition (\ref{norm2}) we find
\begin{equation}
  \Cov(\hat w; z, z') = \frac{c}{\rho} \frac{ \int \diff z'' \,
  p_z(z'') W_z(z, z'') W_z(z', z'') }{ \int \diff^2 \theta'' \, \bigl|
  \gamma(\vec\theta'') \bigr|^2 } \; .  \label{Cov(w)}
\end{equation}
In the case $z = z'$, this equation has a simple interpretation. Let
us call $N_W(z)$ the number of galaxies in the whole field for which
$W_z \bigl( z, z^{(n)} \bigr)$ is significantly different from
zero. Then the variance of $\hat w(z)$ is given simply by $c/\bigl[
N_W (z) \bigl\langle |\gamma|^2 \bigr\rangle \bigr]$, where
$\bigl\langle |\gamma|^2 \bigr\rangle$ is the mean value of the shear
on the field. The covariance of $\hat w$ is in general proportional to
$1/\bigl\langle |\gamma|^2 \bigr\rangle$, and thus strong clusters
should be preferred. However, as discussed before, for strong clusters
the sheet invariance can generate serious problems in the determination
of $w(z)$. This simple reasoning also clarifies the behaviour of
$\Cov(\hat w; z, z')$ as a function of the weight $W_z(z, z')$
used. Since the expected variance of $\hat w(z)$ is proportional to
$1/N_W (z)$, broad functions $W_z(z, z')$ should be preferred. However
it is clear that the coherence length of $\hat w$, i.e.\ the maximum
value of $|z - z'|$ that makes $\Cov(\hat w; z, z')$ significantly
different from zero, is given by $N_W (z)$: this would suggest the use
of a function $W_z$ with small $N_W (z)$.

Equation (\ref{Cov(w)}) has been calculated assuming a generic shear
map $\gamma(\vec\theta)$. In practice, the shear used to calculate
$\hat w(z)$ is the estimation $\hat\gamma(\vec\theta)$, and thus
$\gamma(\vec\theta)$ in Eq.~(\ref{Cov(w)}) should be replaced by
$\hat\gamma(\vec\theta)$. 

\subsubsection{Role of the global scaling invariance}

So far no assumption on $w(z)$ has been made. If we consider the
expectation value of $\hat \gamma$, we see that the {\it assumed\/}
cosmological weight enters through the quantity $K$. We have already
observed that indeed this quantity is related to the global scaling
invariance. In contrast to the expression for $\langle \hat\gamma
\rangle$, the expected covariance of $\hat\gamma$, given by
Eq.~(\ref{Cov(gamma)}), has a more complicated dependence on
$w(z)$. When $K = 1$ the measured $\langle \hat w \rangle(z)$ is
closest to the true $w_0(z)$ (see Eq.~(\ref{<w>})). From
Eq.~(\ref{Cov(gamma)}) we see that the minimum of $\Cov(\hat\gamma)$,
under the constraint $K = 1$, is attained when $\int \diff z' \, w(z')
w_0(z')$ is maximum. It is then easy to show, using the Cauchy-Schwarz
inequality, that this occurs when $w(z) = w_0(z)$.

The situation for the covariance of $\hat w$ is different. In fact,
from Eq.~(\ref{Cov(w)}), it is clear that the constant $K$ does not
enter explicitly in the expression of this covariance. This property
will play an important role in the statistical analysis of Appendix~D.

In conclusion, the global scaling invariance, represented by the
quantity $K$, has different effects on the various quantities. This
quantity enters directly in the expressions of the expected values of
$\hat\gamma$ and $\hat w$, while the expression for $\Cov(\hat w)$ is
not affected. However, the expected values of the shear and of the
cosmological weight differ for an important characteristic. In fact,
if the global scaling invariance could be broken, $\gamma$ would be
estimated also without the knowledge of $w(z)$ (see
Eq.~(\ref{<gamma>})), while $w(z)$ cannot be calculated without the
knowledge of $\gamma$. Therefore, the covariance of $\hat\gamma$,
defined as $\bigl\langle \bigl( \hat\gamma(\vec\theta) - \langle
\hat\gamma \rangle (\vec\theta) \bigr) \bigl( \hat\gamma(\vec\theta')
- \langle \hat\gamma \rangle (\vec\theta') \bigr) \bigr\rangle$, can
be identified with the error $\bigl\langle \bigl(
\hat\gamma(\vec\theta) - \gamma_0(\vec\theta) \bigr) \bigl(
\hat\gamma(\vec\theta') - \gamma_0 (\vec\theta') \bigr) \bigr\rangle$
even if the cosmological weight used is not a good representation of
$w_0$. The same, of course, is not true for $w$, since there is no
relation between $\langle w \rangle$ and $w_0$, if the shear map used
differs significantly from $\gamma_0$.

\subsection{General case}

Calculations for the general case are {\it exceedingly\/}
difficult. If the lens is not too strong (say, $|g| < 0.5$) we might
argue that a first approximation could be given by the results of the
previous section. A posteriori, numerical simulations (described in
the main text) confirm the applicability of the weak lensing analysis
to relatively strong lenses. On the other hand, if the lens is not
weak and the sheet invariance cannot be broken, we have already shown
that the method described here is difficult to apply.


\section{Extraction of cosmological information}

We now have all the tools ready, in order to proceed from the
estimation of $w(z)$ to the desired constraints on the cosmological
parameters. In order to describe practically this function, we start
by {\it discretizing\/} the problem in the redshift variable, which so
far, for simplicity, has been kept continuous. We consider a grid of
points $\{ z_i \}$, with $i = 1, \dots, N_z$. We then introduce the
notation
\begin{eqnarray}
&& w_i = w(z_i) \; ,\\
&& \langle \hat w_i \rangle = \langle \hat w \rangle(z_i) \; ,\\
&& Z_{ij} = \Cov(\hat w; z_i, z_j) \; ,
\end{eqnarray}
for the cosmological weight function, its expectation value, and its
covariance on such a grid. At least for the weak lensing limit, the
expression for $\langle \hat w_i \rangle$ and $Z_{ij}$ can now be
taken to be known (see Eqs.~(\ref{<w>}) and (\ref{Cov(w)})). A similar
discretization is performed in the $\vec\theta$-space. The least
squares method for the extraction of the cosmological information
focuses on the study of the {\it chi-square\/} function:
\begin{equation}
  \chi^2(\xi) = \bigl( k \langle \hat w_i
  \rangle - \hat w_i \bigr) \bigl( Z^{-1} \bigr)_{ij} \bigl( k \langle
  \hat w_j \rangle - \hat w_j \bigr) \; , \label{chi^2}
\end{equation}
with $\xi = (\Omega, \Omega_\Lambda, k)$ a point in the parameter
space $\Xi$. The ambiguity associated with the global scaling
invariance is recognized and taken into account by means of the scale
$k$ which is treated as a free parameter in addition to the physical
parameters $\Omega$ and $\Omega_\Lambda$ that we would like to
constrain. Note that $\Omega$ and $\Omega_\Lambda$ enter in a
complicated, non-linear manner, through the convolution of $w_0$ in
the expression for $\langle \hat w \rangle$. The study of
$\chi^2(\xi)$ can be addressed with three different goals.

\subsection{Point estimation}

Here one is interested in determining the most probable values of the
cosmological parameters. These are the values $\hat\xi$ that minimize
$\chi^2$, obtained by solving the set of equations
\begin{equation}
  \left. \frac{\partial \chi^2}{\partial \xi} \right|_{\hat \xi} = 0 \;
  .
\end{equation}
This could be completed with a determination of the related errors.

Here we should recall that the least squares estimation has the
following properties (see Eadie {\it et al}. 1971): (i) If the
expectation values $\langle \hat w_i \rangle$ have been determined
correctly, then the estimation is {\it consistent}, i.e.\ it leads to
the true cosmological parameters in the limit $N \rightarrow
\infty$. (ii) If the measured weights $\hat w_i$ follow a Gaussian
distribution, then the least squares method is equivalent to the
maximum likelihood method. This occurs when the number of galaxies is
sufficiently high ($N_W(z) \gtrsim 10$). (iii) If the mean $\langle
\hat w \rangle(z)$ is a linear function of the parameters, or if a
linear approximation can be used, the Gauss-Markov theorem states that
the least squares method has minimum variance among the {\it
unbiased\/} estimators that are {\it linear functions\/} of the
observed quantities. This last property reassures us of the quality of
the method used. Fortunately, at least for the scale parameter $k$ the
theorem holds.

\subsection{Interval estimation}

The interval estimation aims at identifying the relevant {\it
confidence regions\/} in the parameter space $\Xi$, starting from a
given {\it confidence level}, $\CL < 1$. The region is usually
determined by referring to the chi-square probability distribution,
which depends on the number of degrees of freedom.

In the present case, given the values of the cosmological parameters,
measurements of the cosmological weight will be realized (in the space
of all possible measurements) with some probability distribution. Let
us isolate in such a space a region with the following property: the
probability that a set of measurements $\{ \hat w_i \}$ be realized
inside it is greater than a desired level, the confidence level
$\CL$. Suppose that a change of variables from $\{ w_i \}$ to $\{ v_i
\}$ can be found such that the corresponding region in the new
$v$-space does not depend on the cosmological parameters. Then, given
one set of measurements $\{ \hat w_i \}$, we may associate that region
in $v$-space to the so-called confidence region in parameter space.

Our case requires additional discussion because one of the parameters,
the scale $k$, is non-physical, and we are interested in defining
confidence regions in the $\Omega$-$\Omega_\Lambda$ plane. In order to
do so, we might project the region found in the $\Xi$-space onto the
plane of physical parameters. Since $k$ can be ignored, an efficient
approach is the following. We note that if we minimize $\chi^2$
with respect to one parameter, the new quantity follows asymptotically
($N \rightarrow \infty$) a chi-square distribution with $N_z - 1$
degrees of freedom. Therefore, we may introduce the quantity
\begin{equation}
  \ell_2 (\Omega, \Omega_\Lambda) = \min_k \chi^2(\xi) - \chi^2( \hat
  \xi) \; , \label{ell2}
\end{equation}
which is expected to follow a chi-square distribution with two degrees
of freedom. With this device, the confidence regions can be drawn in a
straightforward way, independently of the number $N_z$ of points in
the redshift grid. For finite values of $N$, $\ell_2$ is not required
to follow strictly a chi-square distribution with two degrees of
freedom; however, $\ell_2$ is always a reasonable statistic.

The method described above turns out to be just an application of the
well-known ``likelihood ratio'' method to our problem (with the global
scaling invariance taken into account). In fact, it can be shown that
the quantity $\ell_2$ can be expressed as
\begin{eqnarray}
  \ell_2(\Omega, \Omega_\Lambda) &=& 2 \sum_{n=1}^N \Bigl[ \ln {\cal
  L} \bigl( \epsilon^{(n)} \bigm| \hat\Omega, \hat\Omega_\Lambda, \hat
  k, \hat\gamma(\vec\theta^{(n)}) \bigr) \nonumber\\ && \qquad {} -
  \ln {\cal L} \bigl( \epsilon^{(n)} \bigm| \Omega, \Omega_\Lambda,
  \hat k, \hat \gamma(\vec\theta^{(n)}) \bigr) \Bigr] \; .
  \label{LL}
\end{eqnarray}
In this expression ${\cal L}$ is the ``likelihood'' for a single
galaxy (see Eadie {\it et al}.\ 1971), i.e.\ the probability of
observing a galaxy with ellipticity $\epsilon^{(n)}$ when the shear is
$\gamma(\vec\theta)$ and when the ``cosmological'' parameters are
$\Omega$, $\Omega_\Lambda$, and $k$.  Then, using the central limit
theorem and the results (\ref{<w>}) and (\ref{Cov(w)}) for the mean
and the covariance of the cosmological weight, the relation between
Eq.~(\ref{ell2}) and (\ref{LL}) follows.

The quantity $\ell_2$ will prove to be useful also as a test of
hypotheses, as explained below.

\subsection{Test of hypotheses}

The third analysis that can be performed on the measured cosmological
weight is the test of hypotheses. In this case a given hypothesis
$h_0$, called ``null hypothesis,'' is tested against an alternative
hypothesis $h_1$. A test is a method to choose between $h_0$ and $h_1$
from a given set of observations. Hence it can be thought as a region
in the space of the observed quantities: if the point representing the
observations lies inside this area, the null hypothesis is true,
otherwise it is false.

All hypotheses that we may be interested in testing would actually be
{\it composite\/} hypotheses, because of the presence of the scale
parameter $k$. Let $\Xi_0$ be the subset of $\Xi$ for which the null
hypothesis is true (for example, $\Omega + \Omega_\Lambda = 1$, with
$k$ arbitrary). Then the alternative hypothesis holds on the set $\Xi
\setminus \Xi_0$. The likelihood ratio is defined as
\begin{equation}
  l = \frac{ \displaystyle{\max_{\xi \in \Xi_0} {\cal L} \bigl(
  \epsilon^{(n)} \bigm| \xi \bigr)} }{ \displaystyle{\max_{\xi \in
  \Xi} {\cal L} \bigl( \epsilon^{(n)} \bigm| \xi \bigr)} } \;
  . \label{LL+1}
\end{equation}
Clearly $0 \le l \le 1$. Then, the likelihood ratio test is given by
the following rule: if $l < l_0$ then choose $h_0$, otherwise choose
$h_1$. The limiting value $l_0$ has to be chosen so that the
probability of ``loss error'' (i.e., the rejection of the null
hypothesis when in reality $h_0$ is true) is small, and the threshold
value may be set, for example, at $5\%$. Such a test is found empirically to
produce useful and significant results.

If we want to test a hypothesis $h_0$ fixing separately the values of
$\Omega$ and $\Omega_\Lambda$ (so that $\Xi_0$ reduces to a straight
line), by combining the discussion of Eq.~(\ref{LL}) with that of
Eq.~(\ref{LL+1}) we see that we can express $\ell_2$ in terms of
$\chi^2$ as
\begin{equation}
  \ell_2 = -2 \ln l = \min_{\xi \in \Xi_0} \chi^2(\xi) - \min_{\xi \in
  \Xi} \chi^2(\xi) \; .
  \label{ell2-b}
\end{equation}
Thus it can be shown that, if the $h_0$ hypothesis is true, the
quantity $(-2 \ln l)$ follows (asymptotically) a chi-square
distribution with two degrees of freedom.

For different types of $h_0$ the quantity $(-2 \ln l)$ follows
(asymptotically) a chi-square distribution with degrees of freedom
equal to the number of parameters fixed by $h_0$.

At this point, the ``level of acceptance'' can be obtained from the
cumulative chi-square distribution for the relevant number of degrees
of freedom. In particular, if we want to test the Einstein-de~Sitter
universe ($\Omega=1$, $\Omega_\Lambda=0$) with $5\%$ significance, we
use the statistic
\begin{equation}
  \ell_2 = \min_k \chi^2(1, 0, k) - \chi^2(\hat\xi) \; .
\end{equation}
The universe will be considered Einstein-de~Sitter if $\ell_2 <
5.991$. If, instead, we want to test the flat hypothesis ($\Omega +
\Omega_\Lambda = 1$) with significance $5\%$, we use the statistic
\begin{equation}
  \ell_1 = \min_{\Omega, k} \chi^2(\Omega, 1-\Omega, k) -
  \chi^2(\hat \xi) \; .
\end{equation}
The universe will be considered to be flat if $\ell_1 < 3.841$.

As noted earlier, when the asymptotic ($N \rightarrow \infty$) limit is
not justified, one should resort to Monte Carlo simulations in order
to set the appropriate ``level of acceptance.''

\end{appendix}

\end{document}